\documentclass[12pt]{article}
\usepackage[dvips]{graphicx}
\usepackage{epsfig}
\usepackage{latexsym,amsmath,amsfonts,amssymb}
\usepackage[latin1]{inputenc}
\usepackage[american]{babel}
\usepackage[dvips]{graphicx}
\usepackage{bbm}
\usepackage{color}
\usepackage[unicode]{hyperref}
\def\im{Invent. Math.}

\def\a{\alpha}
\def\b{\beta}
\def\c{\gamma}
\def\d{\delta}
\def\f{\phi}               
\def\vf{\varphi}  \def\tvf{\tilde{\varphi}}
\def\vp{\varphi}
\def\g{\gamma}
\def\h{\eta}
\def\j{\psi}
\def\k{\kappa}                    
\def\l{\lambda}
\def\m{\mu}
\def\n{\nu}
\def\o{\omega}  \def\w{\omega}

\def\q{\theta}  \def\th{\theta}                  
\def\r{\rho}                                     
\def\s{\sigma}                                   
\def\t{\tau}
\def\u{\upsilon}
\def\x{\xi}
\def\z{\zeta}
\def\pt{\tilde{\varphi}}
\def\tt{\tilde{\theta}}
\def\lab{\label}
\def\6{\partial}
\def\wg{\wedge}
\def\bpsi{\bar{\psi}}
\def\bt{\bar{\theta}}
\def\bvf{\bar{\varphi}}

\DeclareMathOperator{\tr}{tr}

\newcommand{\be}{\begin{equation}}
\newcommand{\ee}{\end{equation}}
\newcommand{\beq}{\begin{equation}}
\newcommand{\eeq}{\end{equation}}
\newcommand{\bea}{\begin{eqnarray}}
\newcommand{\eea}{\end{eqnarray}}

\newcommand{\ba}{\begin{eqnarray}}
\newcommand{\ea}{\end{eqnarray}}

\newcommand{\beqs}{\begin{eqnarray}}
\newcommand{\eeqs}{\end{eqnarray}}
\newcommand{\bal}{\begin{aligned}}
\newcommand{\eal}{\end{aligned}}

\begin{document}
\baselineskip=15.5pt
\pagestyle{plain}
\setcounter{page}{1}


\def\del{{\partial}}
\def\vev#1{\left\langle #1 \right\rangle}
\def\cn{{\cal N}}
\def\co{{\cal O}}
\def\IC{{\mathbb C}}
\def\IR{{\mathbb R}}
\def\IZ{{\mathbb Z}}
\def\RP{{\bf RP}}
\def\CP{{\bf CP}}
\def\Poincare{{Poincar\'e }}
\def\tr{{\rm tr}}
\def\tp{{\tilde \Phi}}

\def\TL{\hfil$\displaystyle{##}$}
\def\TR{$\displaystyle{{}##}$\hfil}
\def\TC{\hfil$\displaystyle{##}$\hfil}
\def\TT{\hbox{##}}
\def\HLINE{\noalign{\vskip1\jot}\hline\noalign{\vskip1\jot}}
\def\seqalign#1#2{\vcenter{\openup1\jot
   \halign{\strut #1\cr #2 \cr}}}
\def\lbldef#1#2{\expandafter\gdef\csname #1\endcsname {#2}}
\def\eqn#1#2{\lbldef{#1}{(\ref{#1})}%
\begin{equation} #2 \label{#1} \end{equation}}
\def\eqalign#1{\vcenter{\openup1\jot
     \halign{\strut\span\TL & \span\TR\cr #1 \cr
    }}}
\def\eno#1{(\ref{#1})}
\def\href#1#2{#2}
\def\half{\frac{1}{2}}

\def\ads{{\it AdS}}
\def\adsp{{\it AdS}$_{p+2}$}
\def\cft{{\it CFT}}

\newcommand{\ber}{\begin{eqnarray}}
\newcommand{\eer}{\end{eqnarray}}

\newcommand{\beqar}{\begin{eqnarray}}
\newcommand{\cN}{{\cal N}}
\newcommand{\cO}{{\cal O}}
\newcommand{\cA}{{\cal A}}
\newcommand{\cT}{{\cal T}}
\newcommand{\cF}{{\cal F}}
\newcommand{\cC}{{\cal C}}
\newcommand{\cR}{{\cal R}}
\newcommand{\cW}{{\cal W}}
\newcommand{\eeqar}{\end{eqnarray}}
\newcommand{\tht}{\thteta}
\newcommand{\lm}{\lambda}\newcommand{\Lm}{\Lambda}


\newcommand{\nonu}{\nonumber}
\newcommand{\oh}{\displaystyle{\frac{1}{2}}}
\newcommand{\dsl}
   {\kern.06em\hbox{\raise.15ex\hbox{$/$}\kern-.56em\hbox{$\partial$}}}
\newcommand{\id}{i\!\!\not\!\partial}
\newcommand{\as}{\not\!\! A}
\newcommand{\ps}{\not\! p}
\newcommand{\ks}{\not\! k}
\newcommand{\D}{{\cal{D}}}
\newcommand{\dv}{d^2x}
\newcommand{\Z}{{\cal Z}}
\newcommand{\N}{{\cal N}}
\newcommand{\Dsl}{\not\!\! D}
\newcommand{\Bsl}{\not\!\! B}
\newcommand{\Psl}{\not\!\! P}
\newcommand{\eeqarr}{\end{eqnarray}}
\newcommand{\ZZ}{{\rm \kern 0.275em Z \kern -0.92em Z}\;}


\def\del{{\delta^{\hbox{\sevenrm B}}}} \def\ex{{\hbox{\rm e}}}
\def\azb{A_{\bar z}} \def\az{A_z} \def\bzb{B_{\bar z}} \def\bz{B_z}
\def\czb{C_{\bar z}} \def\cz{C_z} \def\dzb{D_{\bar z}} \def\dz{D_z}
\def\im{{\hbox{\rm Im}}} \def\mod{{\hbox{\rm mod}}} \def\tr{{\hbox{\rm Tr}}}
\def\ch{{\hbox{\rm ch}}} \def\imp{{\hbox{\sevenrm Im}}}
\def\trp{{\hbox{\sevenrm Tr}}} \def\vol{{\hbox{\rm Vol}}}
\def\rl{\Lambda_{\hbox{\sevenrm R}}} \def\wl{\Lambda_{\hbox{\sevenrm W}}}
\def\fc{{\cal F}_{k+\cox}} \def\vev{vacuum expectation value}
\def\nodiv{\mid{\hbox{\hskip-7.8pt/}}}
\def\ie{{\em i.e.}}
\def\ie{\hbox{\it i.e.}}

\def\CC{{\mathchoice
{\rm C\mkern-8mu\vrule height1.45ex depth-.05ex
width.05em\mkern9mu\kern-.05em}
{\rm C\mkern-8mu\vrule height1.45ex depth-.05ex
width.05em\mkern9mu\kern-.05em}
{\rm C\mkern-8mu\vrule height1ex depth-.07ex
width.035em\mkern9mu\kern-.035em}
{\rm C\mkern-8mu\vrule height.65ex depth-.1ex
width.025em\mkern8mu\kern-.025em}}}

\def\RR{{\rm I\kern-1.6pt {\rm R}}}
\def\NN{{\rm I\!N}}
\def\ZZ{{\rm Z}\kern-3.8pt {\rm Z} \kern2pt}
\def\IB{\relax{\rm I\kern-.18em B}}
\def\ID{\relax{\rm I\kern-.18em D}}
\def\II{\relax{\rm I\kern-.18em I}}
\def\IP{\relax{\rm I\kern-.18em P}}
\newcommand{\CS}{{\scriptstyle {\rm CS}}}
\newcommand{\CSs}{{\scriptscriptstyle {\rm CS}}}
\newcommand{\rc}{\nonumber\\}
\newcommand{\bear}{\begin{eqnarray}}
\newcommand{\eear}{\end{eqnarray}}

\newcommand{\LL}{{\cal L}}

\def\mani{{\cal M}}
\def\calo{{\cal O}}
\def\calb{{\cal B}}
\def\calw{{\cal W}}
\def\calz{{\cal Z}}
\def\cald{{\cal D}}
\def\calc{{\cal C}}
\def\to{\rightarrow}
\def\ele{{\hbox{\sevenrm L}}}
\def\ere{{\hbox{\sevenrm R}}}
\def\zb{{\bar z}}
\def\wb{{\bar w}}
\def\nodiv{\mid{\hbox{\hskip-7.8pt/}}}
\def\menos{\hbox{\hskip-2.9pt}}
\def\dr{\dot R_}
\def\drr{\dot r_}
\def\ds{\dot s_}
\def\da{\dot A_}
\def\dga{\dot \gamma_}
\def\ga{\gamma_}
\def\dal{\dot\alpha_}
\def\al{\alpha_}
\def\cl{{closed}}
\def\cls{{closing}}
\def\vev{vacuum expectation value}
\def\tr{{\rm Tr}}
\def\to{\rightarrow}
\def\too{\longrightarrow}


\def\a{\alpha}
\def\b{\beta}
\def\c{\gamma}
\def\d{\delta}
\def\e{\epsilon}           
\def\F{\Phi}
\def\f{\phi}               
\def\vf{\varphi}  \def\tvf{\tilde{\varphi}}
\def\vp{\varphi}
\def\g{\gamma}
\def\h{\eta}
\def\j{\psi}
\def\k{\kappa}                    
\def\l{\lambda}
\def\m{\mu}
\def\n{\nu}
\def\o{\omega}  \def\w{\omega}
\def\q{\theta}  \def\th{\theta}                  
\def\r{\rho}                                     
\def\s{\sigma}                                   
\def\t{\tau}
\def\u{\upsilon}
\def\x{\xi}
\def\X{\Xi}
\def\z{\zeta}
\def\pt{\tilde{\varphi}}
\def\tt{\tilde{\theta}}
\def\lab{\label}
\def\6{\partial}
\def\wg{\wedge}
\def\atanh{{\rm arctanh}}
\def\bpsi{\bar{\psi}}
\def\bt{\bar{\theta}}
\def\bvf{\bar{\varphi}}

%

\newfont{\namefont}{cmr10}
\newfont{\addfont}{cmti7 scaled 1440}
\newfont{\boldmathfont}{cmbx10}
\newfont{\headfontb}{cmbx10 scaled 1728}
\newcommand{\re}{\,\mathbb{R}\mbox{e}\,}
\newcommand{\hyph}[1]{$#1$\nobreakdash-\hspace{0pt}}
\providecommand{\abs}[1]{\lvert#1\rvert}
\newcommand{\Nugual}[1]{$\mathcal{N}= #1 $}
\newcommand{\sub}[2]{#1_\text{#2}}
\newcommand{\partfrac}[2]{\frac{\partial #1}{\partial #2}}
\newcommand{\bsp}[1]{\begin{equation} \begin{split} #1 \end{split} \end{equation}}
\newcommand{\calF}{\mathcal{F}}
\newcommand{\calO}{\mathcal{O}}
\newcommand{\calM}{\mathcal{M}}
\newcommand{\calV}{\mathcal{V}}
\newcommand{\bbZ}{\mathbb{Z}}
\newcommand{\bbC}{\mathbb{C}}
\newcommand{\cK}{{\cal K}}

\newcommand{\Thq}{\Theta\left(\r-\r_q\right)}
\newcommand{\Dq}{\d\left(\r-\r_q\right)}
\newcommand{\kten}{\kappa^2_{\left(10\right)}}
\newcommand{\pbi}[1]{\imath^*\left(#1\right)}
\newcommand{\ho}{\hat{\omega}}
\newcommand{\tth}{\tilde{\th}}
\newcommand{\tf}{\tilde{\f}}
\newcommand{\tj}{\tilde{\j}}
\newcommand{\tw}{\tilde{\omega}}
\newcommand{\tz}{\tilde{z}}
\newcommand{\prj}[2]{(\partial_r{#1})(\partial_{\j}{#2})-(\partial_r{#2})(\partial_{\j}{#1})}
\def\atanh{{\rm arctanh}}
\def\sech{{\rm sech}}
\def\csch{{\rm csch}}
\allowdisplaybreaks[1]

\def\red{\textcolor[rgb]{0.98,0.00,0.00}}

\numberwithin{equation}{section}

\newcommand{\Tr}{\mbox{Tr}}    


%
\renewcommand{\theequation}{{\rm\thesection.\arabic{equation}}}
\begin{titlepage}
\rightline{MAD-TH-13-05}
\vspace{0.1in}

\begin{center}
\Large \bf  Dualising the Baryonic Branch:\\
 Dynamic $SU(2)$ \& confining backgrounds in IIA
\end{center}
\vskip 0.2truein
\begin{center}
J\'er\^ome Gaillard$^{a,}$\footnote{jgaillard@wisc.edu}, 
Niall T. Macpherson $^{b,}$\footnote{pymacpherson@swansea.ac.uk},
Carlos N\'u\~nez$^{b,c,}$\footnote{c.nunez@swansea.ac.uk} and 
Daniel C. Thompson$^{d,}$\footnote{dthompson@tena4.vub.ac.be} 
\vskip 0.2truein
 \vskip 4mm
{\it $a$: Physics Department.
University of Wisconsin-Madison
1150 University Avenue\\
Madison, WI 53706-1390. USA}

\vspace{0.2in}
{\it $b$: Department of Physics, Swansea University\\
 Singleton Park, Swansea SA2 8PP, United Kingdom.}

\vspace{0.2in}
{\it $c$: Albert Einstein Minerva Center, Weizmann Institute of Science\\
 Rehovot 76100, Israel.}

\vspace{0.2in}
{\it $d$: Theoretische Natuurkunde, Vrije Universiteit Brussel,
and The International Solvay Institutes,
Pleinlaan 2, B-1050 Brussels, Belgium.
}
\vskip 5mm

\vspace{0.2in}
\end{center}
\vspace{0.2in}
\centerline{{\bf Abstract}}
In this paper we construct and examine new supersymmetric solutions of massive IIA supergravity that are obtained using non-Abelian T-duality applied to the Baryonic Branch of the  Klebanov-Strassler background.  The geometries display $SU(2)$ structure which we show  
flows from static in the UV to dynamical in the IR. Confinement and symmetry breaking are given a geometrical interpretation by this change of structure. Various field theory observables are studied, suggesting  possible ways to break conformality and flow in ${\cal N}=1~T_N$  and related field theories.
\smallskip
\end{titlepage}
\setcounter{footnote}{0}

\tableofcontents

\setcounter{footnote}{0}
\renewcommand{\theequation}{{\rm\thesection.\arabic{equation}}}

 \newpage 
\section{Introduction And General Idea Of This Paper}

 The notion of duality is of course quite old,  going back to well-known examples like the Maxwell equations in vacuum.  
 The true power of the idea became clear around 1940 with the Kramers-Wannier \cite{Kramers}  duality of the Ising model.  In more recent times dualities have continued to be a driver of theoretrical progress with examples including  Bosonisation \cite{Coleman:1974bu}, Montonen-Olive
duality \cite{Montonen:1977sn}, S and T-dualities, Seiberg-Witten duality \cite{Seiberg:1994rs}, Seiberg duality  \cite{Seiberg:1994pq} 
and more general String dualities (U dualities). The duality conjectured by Maldacena \cite{Maldacena:1997re}, 
also called AdS/CFT or Gauge-Strings duality, 
is arguably the most powerful, widely applicable and conceptually 
deep duality of all known at present. All these dualities present common features: the degrees of 
freedom on both sides of the 
dual descriptions are in principle quite different; a 
strongly coupled (highly fluctuating) 
description of the system is characteristically
mapped into a weakly coupled (semiclassical) one, in the same vein
a phenomena that is `local' in one set of variables becomes `non-local'
in the other (as exemplified by order-disorder operators and their typical
`uncertainty' relations),
global symmetries are common to both dual descriptions, etc.

In this paper, we will mostly work with two dualities, 
the one conjectured by 
Maldacena and its extensions (see the papers
\cite{Itzhaki:1998dd} for a sample of representative work and reviews)   
together with what is called `non-Abelian T-duality' 
\cite{de la Ossa:1992vc}. 
We will  use non-Abelian T-duality as 
a technique to generate new solutions 
to the equations of motion of Type II supergravity.  Following the implementation of non-Abelian T-duality as a solution generating technique of RR backgrounds in \cite{Sfetsos:2010uq}, there have been a number of recent developments in the use of non-Abelian duality, see \cite{Lozano:2011kb}-\cite{Lozano:2013oma}. 
We will make use of many technical tools developed in these various papers.
 
We will consider  
backgrounds of Type II Supergravity that have a well understood 
(strongly coupled) 
field theory dual;
we will then study the effect of this
generating technique on the background. 
This  will lead us to the construction of new solutions of 
ten-dimensional Supergravity and, as advocated in
\cite{Itsios:2013wd},
 we will use these new backgrounds 
to define new field 
theories at strong coupling. All of our backgrounds will be smooth and
minimal supersymmetry in four dimensions (four supercharges) 
will be preserved. These new solutions 
will admit a description in terms of G-structures and we will
explain how certain field theoretical phenonomena, 
like confinement and symmetry 
breaking are encoded in generic changes of the G-structure.

The system on which we will focus our study  
is the Baryonic Branch of the Klebanov-Strassler field theory
\cite{Klebanov:2000hb}, \cite{Gubser:2004qj}, \cite{Butti:2004pk}.
This is perhaps, among the minimally SUSY examples known at the moment, 
the one that
better passed test of the correspondence between geometry and 
(strongly coupled) field 
theoretical aspects. Besides, the Baryonic Branch field theory 
and geometry unifies 
the original Klebanov-Strassler system and the system 
of five branes wrapping a two 
cycle inside the resolved conifold 
\cite{Maldacena:2000yy}. Field theoretically, this unification 
can be thought as a Higgs-like mechanism and a particular limit where 
an accidental symmetry appears. See the papers in  \cite{Maldacena:2009mw} 
for different geometric and physical
aspects of this connection.

In this work, we will perform an $SU(2)$ non-Abelian 
T-duality on the Baryonic Branch geometry. 
This is a geometry described by an $SU(3)$-structure. 
All features of the geometry are 
characterised by a couple of forms $J_2, \Omega_3$ that also encode 
many aspects of the strongly coupled dual field theory. 
Using non-abelian T-duality, 
we will obtain a new background in 
Massive Type IIA Supergravity. The G-structure will  
change to what is called $SU(2)$-structure, 
characterised by forms $j_2, w_1, v_1,\omega_2$. 
The $SU(2)$-structure will transition from being {\it static}
in the large radius region of the geometry 
(corresponding to high energies in the dual field theory) 
to being {\it dynamical} once the small radius region of 
the geometry is considered. 
Hence,  the phenomena of confinement and 
symmetry breaking are given a  geometric description by the change in  
$SU(2)$-structure  from static to dynamical. 
 
The action of non-Abelian T-duality on the G-structures has been studied in many backgrounds which we take the opportunity  to summarise in the table below.\footnote{The details of the case of $Y^{p,q}$ are to appear in \cite{Ypqtoappear} and a detailed study of the D6 branes on $S^3$ will appear in \cite{NCE}.}   

\begin{center}
  \begin{tabular}{ |l | c | r| }
   \hline
  	\textbf{Seed Solution} & \textbf{Seed Structure} & \textbf{Dual Structure}\\\hline
    \hline
    Klebanov-Witten & $SU(3)$ & Orthogonal $SU(2)$ \\ \hline
    Klebanov-Tseytlin & $SU(3)$ & Orthogonal $SU(2)$ \\ \hline
    $ Y^{p,q}  $ &  $SU(3)$  &Orthogonal $SU(2)$ \\ \hline
    Klebanov-Strassler & $SU(3)$ & Dynamical $SU(2)$  \\\hline
    KS Baryonic Branch & $SU(3)$ & Dynamical $SU(2)$  \\\hline
    Wrapped D5's on $S^2$ & $SU(3)$ & Dynamical $SU(2)$\\ \hline
    Wrapped D6's on $S^3$ & $SU(3)$ & Dynamical $SU(2)$\\ \hline
		Wrapped D5's on $S^3$ & $G_2$ & Dynamical $SU(3)$\\ \hline
    \hline
  \end{tabular}
\end{center}

The contents of this paper are organised as follows. 
In Section \ref{sectionofbaryonicbranch}
 we will briefly summarise the original background and field theory  
corresponding to the Baryonic Branch of the Klebanov-Strassler field theory 
(the {\it seed} background/field theory pair 
on which we will apply our generating technique). In Section \ref{NATDBB}
we will present explictly the new solution. 
 
In Section \ref{G-structures}, we will organise all the previous information
using the language of G-structures. 
This will lead to a compact way of writing things, that
can be very useful for other studies. 
We will study how the dynamical or static 
character of the G-structure depends on the
field theoretic low energy dynamics captured by the original solution.
In Section \ref{correspondenceqft}, we will discuss  different
aspects of the field theory dual to our new backgrounds. 
We close the paper with a list 
of possible future problems and conclusions. 
A number of technical and useful appendixes complement
our presentation.

\section{Generalities on the Baryonic Branch}\label{sectionofbaryonicbranch}

The Klebanov-Strassler field theory is a two-group  quiver with bifundamental
matter, charged under a global symmetry of the form
$SU(2)\times SU(2)\times U(1)_R\times U(1)_B$. The ranks of the gauge groups 
are ($N, N+M$) and the bifundamental matter $A_1, A_2, B_1, B_2$ 
self-interact via a superpotential of the form
${\cal W}\sim ABAB$. For a very clear explanation 
of many of the details of this quantum field theory,
see \cite{Strassler:2005qs}, \cite{Dymarsky:2005xt}. 
One detail that will be crucial to our present
work is the fact that the so called 
`duality cascade', a succesion of Seiberg dualities,
ends in a situation where the quantum field theory may choose to
develop VEV's for the Baryon and anti-Baryon operators.  

In the last step of the duality cascade the gauge group is $SU(M)\times SU(2M)$.  This theory has mesons ${\cal M}  = (A_a)^{ \a}_{ i} (B_b)^i_{\b}$ and also baryonic operators
\cite{Gubser:2004qj}
\beq
{\cal B}=\epsilon_{\alpha_1....\alpha_{2M}} 
(A_1)_{1}^{\alpha_1}
(A_1)_{2}^{\alpha_2} ....
(A_1)_{M-1}^{\alpha_{M-1}}
(A_1)_{M}^{\alpha_M}\times 
(A_2)_{1}^{\alpha_{M+1}}
(A_2)_{2}^{\alpha_{M+2}}\times ....
(A_2)_{M-1}^{\alpha_{2M-1}}
(A_2)_{M}^{\alpha_{2M}}
\eeq
and similar for $\tilde{{\cal B}}$ made out of $(B_{i})^{a}_{l}$ fields. One can see that both baryons and anti-baryons are neutral under $SU(2)\times SU(2)$ transformations.    

The moduli space consists of two branches - the {\it{ mesonic}} and the {\it{baryonic}}  \cite{Dymarsky:2005xt}.  On the mesonic branch the baryons are zero (${\cal B} =\tilde  {\cal B}= 0$)   and the mesons satisfy $ \det {\cal M} =\Lambda^{4M}$. The non-perturbative contribution to the superpotential means that the associated moduli space can be identified with a symmetric product of the deformed conifold. On the baryonic branch the mesons are zero (${\cal M}=0$) but the baryons acquire expectation values,
\beq
{\cal B}=i \xi \Lambda^{2M},\;\;\; \tilde{\cal{B}}= \frac{i}{\xi}\Lambda^{2M} \ ,
\eeq
where $\Lambda $ is the strong coupling scale of the group $SU(2M)$.
Notice that both VEV's are equal only if $\xi=1$. This corresponds
to a  $\mathbb{Z}_2$-symmetric point, represented by the exact solution
in \cite{Klebanov:2000hb}. 

On this baryonic branch the $U(1)_B$ symmetry 
is spontaneously broken and the associated massless 
(pseudo-scalar) Goldstone mode corresponds to the phase of $\xi$.  By supersymmetry this Goldstone lives in a chiral multiplet and comes along with scalar partner,  the saxion, which corresponds to changing the modulus of $\xi$. As discussed in \cite{Dymarsky:2005xt},  the VEV of the operator,
\beq
{\cal U}= \Tr [A_i  A^\dag_i - B_j B^\dag_j] \ ,
\eeq
 which contains the $U(1)_B$ current $J_\mu$ as its $\theta  \sigma^\mu \bar \theta$ component,  encodes the motion along the baryonic branch (the different values of $\xi$) according to  
\beq
\langle {\cal U} \rangle \sim M \Lambda^2 \ln | \xi | \ . 
\eeq

%
%
%
Let us focus on the situation where the field theory 
chooses to move to the purely baryonic branch.
In this case, there is a smooth solution of 
the equations of motion of Type IIB supergravity,
that describes the strong dynamics of this field theory, 
including the spontaneous breaking
of the $U(1)_B$ symmetry \cite{Gubser:2004qj}, \cite{Butti:2004pk}. 
In the notation 
that we will adopt in this work, such background can 
be written compactly by introducing
the (string frame) vielbein basis,
\begin{align}
	e^{x^i}			&= e^{\frac{\Phi}{2}}
\hat{h}^{-\frac{1}{4}} dx^i    
\,,\;\;\;
	e^{\r}	 		=  e^{\frac{\Phi}{2}+k} 
\hat{h}^{\frac{1}{4}}d\r  
\,,\;\;\;
	e^{\theta}	=  e^{\frac{\Phi}{2}+h} 
\hat{h}^{\frac{1}{4}}    d\theta  
\,,\;\;\;
	 e^{\varphi}= e^{\frac{\Phi}{2}+h} 
\hat{h}^{\frac{1}{4}} 	\sin\theta \,
d\varphi\,    ,\nonumber\\
	e^{1}				&=  \frac{1}{2}e^{\frac{\Phi}{2}+g} 
\hat{h}^{\frac{1}{4}}	(\tilde{\omega}_1 +a\, d\theta)\,  ,\qquad\qquad
	e^{2}				=		\frac{1}{2}
e^{\frac{\Phi}{2}+g} 
\hat{h}^{\frac{1}{4}}	(\tilde{\omega}_2 	
-a\,\sin\theta\, d\varphi)   
\,,\nonumber\\ 
	e^{3}				&= \frac{1}{2}e^{\frac{\Phi}{2}+k} 
\hat{h}^{\frac{1}{4}}	(\tilde{\omega}_3 +\cos\theta\, d\varphi)\,.
	\label{vielbeinafter}
\end{align}
Where $\tilde\omega_i$ are the left invariant forms of $SU(2)$.
The metric, RR and NSNS fields are
\bea
& & ds^2= \sum_{i=1}^{10} (e^{i})^2\,,\nonumber\\
& & F_3= \frac{e^{-\frac{3}{2}\Phi}}{\hat{h}^{3/4}}
\Big[f_1 e^{123}+ f_2 e^{\theta\varphi 3}
+ f_3(e^{\theta23}+ e^{\varphi 13})+ 
f_4(e^{\r 1\theta}+ e^{\r\varphi 2})   \Big]\,,\nonumber\\
& & B_2= \k\, \frac{e^{\Phi}}{\hat{h}^{1/2}}\Big[e^{\r 3}-\cos\alpha
(e^{\theta\varphi}+ e^{12})-\sin\alpha(e^{\theta2}+ e^{\varphi 1})   
\Big]\,,\nonumber\\
& & H_3=-\k\, \frac{e^{\frac{1}{2}\Phi}}{\hat{h}^{3/4}}
\Big[-f_1 e^{\theta\varphi \r} - f_2 e^{\r 12}
- f_3(e^{\theta 2\r}+ e^{\varphi 1\r})+
f_4(e^{ 1\theta 3}+ e^{\varphi 2 3})   \Big]\,,\nonumber\\
& & C_4= -\k\,\frac{e^{2\Phi}}{\hat{h}} 
dx^0\wedge dx^1 \wedge dx^2\wedge dx^3\,,  \nonumber\\
& & F_5= \k\, e^{-\frac{5}{2}\Phi -k}\hat{h}^{\frac{3}{4}}
\partial_\r \left(\frac{e^{2\Phi}}{\hat{h}}\right)
\Big[e^{\theta\varphi 123 }- e^{x^0 x^1 x^2 x^3 \r}   \Big]\,.
\label{configurationfinal}
\eea
We have defined
\beq
\cos\alpha= 
\frac{\cosh(2\r)-a}{\sinh(2\r)}\,,\qquad \sin\alpha= 
-\frac{2e^{h-g}}{\sinh(2\r)}\,,\qquad\qquad\hat{h}=1-\k^2 e^{2\Phi}\,,
\label{cosinh}
\eeq
where $\k$ is a constant that we 
will choose to be $\k= e^{-\Phi(\infty)}$. The functions are,
\beq\begin{split}\label{eq: fidef}
&f_1=-2 N_c e^{-k-2g}\,,\qquad\qquad f_2= \frac{N_c}{2}e^{-k-2h}(a^2 -
2 a b +1 )\,,\\
&f_3= N_ce^{-k-h-g}(a-b)\,,\qquad f_4=\frac{N_c}{2}e^{-k-h-g}b' \,.
\end{split}\eeq
The system has a radial coordinate $\r$, on which $(a,b,\Phi,g,h,k)$ depend,
and we have set $\alpha^{\prime}g_s=1$.
The background is then determined by 
solving the equations of motion for the 
functions $(a,b,\Phi,g,h,k)$. 
A system of BPS equations is derived. These non-linear and coupled 
first-order equations
can be arranged in a convenient form, by rewriting the
functions of the background in terms of a combination of them,
that decouples the equations (as 
explained in
\cite{HoyosBadajoz:2008fw}-\cite{Casero:2007jj}).
We will not go over these in the present paper. Enough will be for us to
state that the whole dynamics of the string background is controlled by a 
single function $P(\rho)$, subject to a  second order
non-linear and ordinary differential equation.   This function $P(\rho)$ can be determined numerically and  has IR and UV behaviors 
\begin{equation}
\begin{aligned}
UV: \quad P = e^{4\rho/3} \left[ c_+  + \dots \right] \ , \qquad \rho \rightarrow \infty \ , \\
IR: \quad P  = h_1 \rho + {\cal O}(\rho^3) \ , \qquad \rho \rightarrow 0  \ . 
\end{aligned} 
\end{equation}
There is only one independent parameter, $c_+>0$ (the constant $h_1$ is determined by $c_+$)  and  it is this parameter that can be identified with the Baryonic expectation value 
\beq
 {\cal U} \sim \frac{1}{c_+ } \ . 
\eeq
It is convenient to define a dimensionless quantity $\lambda = 2^{2/3} c_+ \epsilon^{-4/3}$ where $\epsilon$ may be identified with the conifold deformation.
See the paper 
\cite{Conde:2011aa} 
for a good account
of the logic and technical details.

\subsection{$SU(3)$ structure of the Baryonic Branch}

The Supergravity background above is characterised 
by what is called an $SU(3)$
structure. That is, there exists a couple of forms $\hat{J}_2$ and $\hat{\Omega}_3$, in terms of which  the BPS equations, the fluxes and various other quantities characterising the space can be written.

The observation of \cite{Gaillard:2010qg}, 
it that the forms $\hat{J}, \hat{\Omega}$, describing the full Baryonic Branch can be obtained 
from the simpler ones describing a set of D5 branes wrapping the
two cycle of the resolved conifold.  We will not repeat the details
of the derivation here, but we quote  the results to the extent that we will find useful.

In general,   an  $SU(3)$ structure solution can be described by the following pure spinors in type-IIB 
\cite{Martucci:2005ht},
\beq\label{eq: rotatedstructure}
\Psi_+ = - e^{i \zeta(r)} \frac{e^{A}}{8}e^{-i \hat{J}},
~~~ \Psi_-=- i\frac{e^{A}}{8}\hat{\Omega}_{hol}.
\eeq
Where $e^{2A}$ is the warp factor of the metric. Let us define
\beq
e^{i \zeta(r)}= \mathcal{C}+ i \mathcal{S}
\eeq
where $\mathcal{C}^2+\mathcal{S}^2=1$.
It is possible to show that for 
zero axion field, that is
$F_1=0$, SUSY requires the following equalities to hold 
(these are the BPS equations previously mentioned)
\beq\label{eq: RotSU3}
\begin{split}
& d\big(e^{-\Phi}\mathcal{S}\big)=0, \;\;\;\; d\big(e^{2A-\Phi} \mathcal{C}\big)=0,\\
&d\big(e^{3A-\Phi}\hat{\Omega}_{hol}\big)=0, \;\;\;\; 
d\big(e^{4A-2\Phi} \hat{J}\wedge \hat{J}\big)=0.
\end{split}
\eeq
The fluxes are determined as
\beq
\begin{split}\label{eq RotSU3Fluxes}
  B_2=\frac{\mathcal{S}}{\mathcal{C}} \hat{J} \ , \quad  \frac{1}{\mathcal{C}^2} d\big(e^{2A} \hat{J}\big)= e^{4A}\star_6 F_3 \ , \quad   d\big( e^{4A-\Phi}\mathcal{S}\big)=- e^{4A}\star_6 F_5 \ . 
\end{split}
\eeq
 
The system of $N_c$ D5 branes wrapped on the resolved conifold is supported by just $F_3$ flux and   is a solution to these equations when $\mathcal{S} = 0$.  The (string-frame) frame fields that describe this geometry can be obtained from  those of eq.(\ref{vielbeinafter}) by setting $\hat{h}=1$.  In terms of these, the   $J_2, \Omega_3$ ( denoted without hats to distinguish them from those of the Baryonic Branch)  are given by
\begin{equation}\label{eq: JJSU3}
\begin{aligned}
J &= e^{r3}+(\cos\alpha e^{\varphi}+\sin\alpha e^{2})\wedge 
e^{\theta}+(\cos\alpha e^{2}-\sin\alpha e^{\varphi})\wedge e^{1}\,\,, \\[0.1in]
\Omega_{hol} &=\, \big( e^r + i\, e^3 \big) \wedge
\big( (\cos \alpha \,e^{\varphi} + \sin \alpha\, e^2 ) + i\, e^{\theta} \big) 
\wedge \big( (-\sin \alpha\, 
e^{\varphi} + \cos \alpha\, e^2 ) + i\, e^1 \big)\,\, ,
\end{aligned}
\end{equation}
which  obey the relations
$
J\wedge\Omega_{hol}=0,~~~J\wedge J\wedge J= \frac{3 i}{4} 
\Omega_{hol}\wedge \bar{\Omega}_{hol}.
$
The BPS equations for the functions $h,g,k,a,b,\Phi$ and 
the RR three-form flux, are
\beq\label{eq: SU3SUSY}
\begin{split}
& d(J\wedge J)=0,\;\;\;\; d(e^{\Phi/2}\Omega_{hol})=0\\
&d(e^{\Phi} J)+ e^{2\Phi}\star_6 F_3=0.
\end{split}
\eeq
Then the results of   \cite{Gaillard:2010qg}  show that the $\hat{J}, \hat{\Omega}$ of the full Baryonic Branch solution are obtained by introducing a non-zero phase or rotation parameter\footnote{This parameter can also be understood in terms of the boost parameter that enters in the duality chain that relates the wrapped brane geometries to the Baryonic branch \cite{Maldacena:2009mw}  .}  $\zeta(r)$  in to  \eqref{eq: rotatedstructure} and defining:  
\beq
\hat{J}= \mathcal{C} J,~~~~\hat{\Omega}_{hol}= \mathcal{C}^{3/2}\Omega_{hol},
~~ e^{2A}=\frac{e^{\Phi}}{\sqrt{\mathcal{C}}},~~~ \mathcal{S}= e^{\Phi-\Phi_{\infty}}, 
\eeq
where $e^{2A}$ is the warp factor of the Baryonic Branch solution.   For further details on the geometry and physics implied by this `scaling of forms', we refer the reader to the original papers \cite{Gaillard:2010qg}
and \cite{Maldacena:2009mw}.

\subsection{A useful gauge transformation}
 
Let us   comment on a small subtlety that will be important in what follows.   The above rotation argument makes it quite clear that by sending $\zeta\rightarrow 0$, the geometry becomes that of the wrapped D5 branes.  On the other hand taking $\zeta \rightarrow \frac{\pi}{2}$ accompanied with $\lambda\rightarrow 0$, the geometry becomes that given by Klebanov and Strassler i.e. the $\mathbb{Z}_2$ point of the Baryonic Branch. Taking this limit is slightly delicate.  One finds that $\sin\zeta \rightarrow 1 $ and $\cos \zeta \rightarrow \frac{1}{\lambda} h_{KS}$  where $h_{KS}$ is the   Klebanov-Strassler warp factor.  Expanding the functions  $(a, b , \Phi , g, h ,k)$ in the large $\lambda$ limit and rescaling Minkowski coordinates $x_i \rightarrow x_i \lambda^{-1}$ one finds that leading term of the metric is independent of $\lambda$ and reproduces the KS geometry.   The limit applied on the NS two form is less trivial, in fact its expansion in inverse powers of $\lambda$ is 
\begin{equation}\label{Blimit}
B_2  = \lambda \frac{ \epsilon^2 \sinh(2 \rho) }{ 2 \sqrt{3} \kappa P_1 \sqrt{P_1'} } d(P_1 (\tilde{\o}_3 + \cos \theta d\varphi ) - B_{KS} + {\cal O} (\lambda^{-1} ) \ . 
\end{equation}
However the form of $P_1$ (the leading contribution of $P(\rho)$ in this expansion) ensures that the pre-factor on the first term in this expression reduces to a constant and one recovers the Klebanov-Strassler NS two form {\em modulo a pure gauge term}.  

In fact it is going to suit our purposes to perform a similar gauge transformation across the whole baryonic branch \eqref{configurationfinal}.  We do this by defining 
\be\label{eq:gaugeseed}
B_2\to B_2 + d(\mathcal{Z(\rho)}(\tilde{\o}_3+\cos\theta d\varphi)),~~~~~\mathcal{Z}=-\frac{1}{2}\int_0^\rho e^{2k(\rho')+\Phi(\rho')}\mathcal{S}(\rho')d\rho'
\ee
In the KS limit this reduces to exactly the gauge transformation required in \eqref{Blimit} and it has the effect of removing certain mixing between the angular directions and the radial direction in the NS two-form.\footnote{This transformation leaves unchanged the gauge coupling defined through the integral of $B_2$ however it is non-vanishing at infinity and so one should exercise appropriate caution.  }  This will greatly simplify matters upon performing a duality transformation.

%
%
\section{Non-Abelian duality on the Baryonic Branch}\label{NATDBB}
In this section, we will present the result for the non-Abelian T-duality
 when applied to one of the $SU(2)$ isometries of the baryonic branch 
background in eq.(\ref{vielbeinafter})-(\ref{configurationfinal}). We extend the results of \cite{Itsios:2013wd} in which the NS sector was established but full details of the RR sector were not provided.\footnote{The results of \cite{Itsios:2013wd} lead at first sight to a geometry that has  a mixing between angular and radial directions.  
This is however a gauge artifact as will be made clear in 
Appendix \ref{detailsBB}.  
By making the gauge transformation \eqref{eq:gaugeseed} to   the seed geometry, as we do here, one removes this mixing.  Alternatively one can perform the following coordinate transformation to the solution presented in \cite{Itsios:2013wd} to obtain the solution presented here:
\be
v^{there}_3 \rightarrow  v_{3}^{here} + \sqrt{2}\mathcal{Z}, 
\ee
} 
We will perform the transformation described in \cite{Itsios:2013wd} to the coordinates  $(\tilde{\theta},\tilde{\varphi}, \psi)$, present in the left-invariant forms 
of $SU(2)$, $\tilde{\omega}_i, i=1,2,3 $ of eq.(\ref{vielbeinafter}).
We will choose a gauge where the new coordinates after the duality will be
$(v_2,v_3,\psi)$.   We presents the results here and refer the reader to the Appendix \ref{detailsBB} for details.

We will start by specifying the vielbeins.
The components 
\begin{align}
	e^{x^i}			&= e^{\frac{\Phi}{2}}
\hat{h}^{-\frac{1}{4}} dx^i    
\,,\;\;\;
	e^{\r}	 		=  e^{\frac{\Phi}{2}+k} 
\hat{h}^{\frac{1}{4}}d\r  
\end{align}
do not change. The vielbeins in the $(\theta,\varphi)$ directions
 are also unchanged by the duality however we find it useful to introduce a 
rotation in $(e^{\theta},e^{\varphi})$ 
such that the dual solution has no explicit $\psi$ dependence.
\beq
e^{\hat{\theta}}= \sqrt{\mathcal{C}}e^{h+\Phi/2} 
\o_1,~~~~ e^{\hat{\varphi}}=\sqrt{\mathcal{C}}e^{h+\Phi/2}\o_2,
\eeq
where we have introduced left invariant $SU(2)$ forms for the angles $\{\theta, \phi,\psi\}$.
The vielbeins in the directions $\hat{1},\hat{2}, \hat{3}$ and NS 2-form potential can be compactly written in terms of the quantities defined as,
\begin{equation}
\begin{aligned}
\mathcal{H}&=\frac{2\sqrt{2} v_3+4 \mathcal{Z}+ e^{2g+\Phi}\mathcal{S}\cos\alpha} {2\sqrt{2}}
  \ ,  \\
  \mathcal{Z}&=- \frac{1}{2}\int^{\rho}_0\mathcal{S}e^{\Phi+2k} d\rho'  \ ,  \\ 
\mu_1&= a e^g \cos\alpha +2e^h \sin\alpha \ .  
\label{definitionsz}
\end{aligned}
\end{equation}
The function $\mathcal{Z}$ was introduced as a gauge transformation to the seed solution already in \eqref{eq:gaugeseed}.
With these, we have
\beq
\begin{split}\label{eq :BBDualVeil}
e^{\hat{1}}&=\frac{e^{g+\Phi/2}}{8 \mathcal{W}}\sqrt{\mathcal{C}}
\bigg[4e^{2k +\Phi}\mathcal{C}\mathcal{H}(a\mathcal{H}\o_1-v_2\o_3)-\sqrt{2}e^{2(g+k+\Phi)}\mathcal{C}^2(dv_2+a\mathcal{H}\o_2)\\
&~~~~~~~~~~~~~~~~~~~~~~~-8\sqrt{2}v_2(v_2dv_2+\mathcal{H} dv_3) +\frac{1}{2} \mu_1 \mathcal{S} e^{g+\Phi}(8 v_2^2\o_2 +e^{2k+\Phi}\mathcal{C}(e^{2g+\Phi}\mathcal{C}\o_2-2\sqrt{2}\mathcal{H}\o_1))\bigg] \ , \\[3 mm]
e^{\hat{2}}&=\frac{e^{g+3\Phi/2+g}}{8 \mathcal{W}}\mathcal{C}^{3/2}
\bigg[4e^{2g}v_2 (dv_3-av_2\o_2)-4\mathcal{H}e^{2k}(dv_2+a \mathcal{H} \o_2) \\
&~~~~~~~~~~~~~~~~~~~~~~-\sqrt{2}\mathcal{C}e^{2k+2g+\Phi}(a\mathcal{H}\o_1-v_2\o_3) +\frac{1}{2} \mu_1 \mathcal{S} e^{g+2k+\Phi}(e^{2g+\Phi}\mathcal{C} \o_1+2\sqrt{2} \mathcal{H} \o_2)\bigg] , \\[3 mm]
e^{\hat{3}}&=\frac{e^{k+\Phi/2}}{8 \mathcal{W}}\sqrt{\mathcal{C}}
\bigg[4\mathcal{C}v_2e^{4g +\Phi}(v_2\o_3-a\mathcal{H}\o_1)-\sqrt{2}\mathcal{C}^2(dv_3-v_2a \o_2)\\
&~~~~~~~~~~~~~~~~~~~~~~~-8\sqrt{2} \mathcal{H}(v_2dv_2+\mathcal{H}dv_3)+e^{g+\Phi}\mu_1v_2\mathcal{S}(\sqrt{2}\mathcal{C} e^{2g+\Phi}\o_1+4\mathcal{H}\o_2)\bigg] \ .
\end{split}
\eeq
We will then have a metric that in terms of these vielbeins reads,
$ds_{st}^2=\sum_{i=1}^{10} (e^{i})^2$.

In terms of these vielbeins, the NS two-form $B_2$ reads,
\beq\label{eq: BBB2}
\begin{split}
\widehat{B}_2  = -& \frac{1}{4v_2}
\bigg(2e^{-h}a(e^gv_2 e^{\hat{\theta}\hat{1}}+e^{k}
\mathcal{H} e^{\hat{\theta}\hat{3}})-4e^{k-g}
\mathcal{H} e^{\hat{1}\hat{3}}+\sqrt{2}
\mathcal{C}e^{g+k+\Phi}e^{\hat{2}\hat{3}}\bigg)+\\
&~~~ \frac{\mathcal{S}}{\mathcal{C}}\bigg[\frac{\mathcal{H}e^k}{2v_2}
\big(2 e^{-g} e^{\hat{1}\hat{3}}-ae^{-h} e^{\hat{\theta}\hat{3}}\big)+
\frac{e^{g+k+\Phi-h}}{4 \sqrt{2} v_2}\mathcal{C} 
\big(\mu_1e^{\hat{\theta}\hat{3}}-2 e^{h}e^{\hat{2} \hat{3}} \big)-\\
& ~~~~~~~~\frac{e^{-h}}{2} \big(2e^{-h-\Phi}\frac{\mathcal{Z}}{\mathcal{S}}+2e^h\cos\alpha-ae^g\sin\alpha \big)e^{\hat{\theta}\hat{\varphi}}-\frac{e^{-h}}{2}(ae^{g} e^{\hat{\theta}\hat{1}}+\mu_1 e^{\hat{\theta}\hat{2}})\bigg].
\end{split}
\eeq
The dual dilaton is given by
\be\label{eq: BBdil}
\widehat{\Phi} =  \Phi - \frac{1}{2} \ln {\cal W}\ ,\;\;\;
\mathcal{ W} = \frac{\mathcal{C}}{8}\bigg( e^{4g+2k+3\Phi}
\mathcal{C}^2 + 8e^{2g+\Phi}v_2^2 + 8e^{2k+\Phi}\mathcal{H}^2\bigg) \ .
\ee
And the RR sector is given by,
\beq
\begin{split}
\vspace{3 mm}
F_0 &= \frac{N_c}{\sqrt{2}}\ , \\[3 mm]
\vspace{3 mm}
F_2 &= -\frac{e^{-\Phi}}{4\mathcal{C}}N_c\bigg[2e^{-2h}\big(1+a^2-2ab\big)
\mathcal{H} e^{\hat{\theta}\hat{\varphi}} + 
e^{-g-h-k}\mathcal{C}\big(a-b\big)
\bigg( \sqrt{2}e^{2g+k+\Phi}\big(e^{\hat{\theta}\hat{1}}-
e^{\hat{\varphi}\hat{2}}\big)+\\
&~~~~~~~~~~~~4e^{k}\mathcal{H}
\big(e^{\hat{\theta}\hat{2}}-e^{\hat{\varphi}\hat{1}}\big)-
4v_2e^{g}e^{\hat{\varphi}\hat{3}}\bigg)-
8e^{-2g}\mathcal{H} e^{\hat{1}\hat{2}}-8e^{-g-k}v_2e^{\hat{2}\hat{3}}
-2e^{-h-k}v_2e^{r\hat{\theta}}\bigg]-\\
&~~~~~~~\frac{\mathcal{S}e^{g-h}}{\sqrt{2}
\mathcal{C}\sin\alpha}\bigg(N_c b+ a(e^{2g}\cos^2\alpha-N_c)
+e^{g+h}\sin2\alpha\bigg)e^{\hat{\theta}\hat{\varphi}} \ ,  \\[3 mm]
F_4&=\frac{e^{-g-h-k-\Phi}}{8\mathcal{C}}N_c
\bigg[\mathcal{C}\big(1+a^2-2ab\big)e^{\hat{\theta}\hat{\varphi}}
\wedge\big(\sqrt{w}e^{2g+k+\Phi-h}
e^{\hat{1}\hat{2}}+4e^{2g-h}e^{\hat{1}\hat{3}}\big)\\
&~~~~~~~~~~~~~~~~~~~~~~~~\mathcal{C}b'e^{r\hat{\theta}}
\wedge\big(4e^{k}\mathcal{H} e^{\hat{1}\hat{3}}-
\sqrt{2}e^{2g+k+\Phi}e^{\hat{2}\hat{3}}\big)-8e^{g}v_2
\big(a-b\big)e^{\hat{\theta}\hat{1}\hat{2}\hat{3}}\\
&~~~~~~~~~~~~~~~~~~~~~~~~e^{r\hat{\varphi}}\wedge 
\big(4e^{g}v_2e^{\hat{1}\hat{2}}-b'e^k(\sqrt{2}e^{2g+\Phi}
e^{\hat{1}\hat{3}}+4\mathcal{H} e^{\hat{2}\hat{3}})\big)\bigg]-\\
&~\frac{2\mathcal{S}e^{-g-h-k-\Phi}}{\mathcal{C}^2 \sin\alpha}
\bigg(a\big(e^{2g}\cos^2\alpha-N_c\big)+\big(N_c b+e^{g+h}
\sin2\alpha\big)\bigg)
\bigg(\mathcal{H} e^{k}e^{\hat{\theta}\hat{\varphi}\hat{1}\hat{2}}
+v_2e^{g} e^{\hat{\theta}\hat{\varphi}\hat{2}\hat{3}}\bigg) \ . 
\end{split}
\eeq

{\bf Warning on  potentially confusing nomenclature}:   The $N_c$ appearing in the above originated as the number of $D5$ branes wrapping the resolved conifold which was then rotated to give the Baryonic Branch and then T-dualised to this solution.   Prior to T-duality, $N_c$ corresponds to the $D5$ charge which is also commonly denoted by $M$ (which we  will also use in section 5 when  we specialised to the Klebanov-Tseytlin geometry).   We hope the reader will not get overly confused by this point. 

\subsection{UV asymptotic behaviour}
Using the semi analytic UV expansions that can be found, for example, in \cite{Conde:2011aa} it is possible to calculate the UV behaviour of the dual metric.  The dual vielbeins at leading order in the UV are given by
\beq
e^{\hat 1}=-\frac{c_+e^{-2\rho/3}(24\rho-3)^{1/4}}{2^{3/4}\sqrt{N_c}(1-2\rho)}\o_1,~~e^{\hat 2}=\frac{c_+e^{-2\rho/3}(24\rho-3)^{1/4}}{2^{3/4}\sqrt{N_c}(1-2\rho)}\o_2,~~e^{\hat 3}= -\frac{2^{3/4} 3^{1/4}}{\sqrt{N_c}(8\rho-1)^{1/4}}dv_3.
\eeq
Thus the dual 3-manifold shrinks as one flows towards the UV, in line with our expectations from abelian T-duality, where big circles are mapped to small circles.

One may worry that this vanishing manifold is a signal of a singularity in the UV, however, an explicit check shows that the curvature invariants: Ricci scalar, 
$R_{\mu\nu}R^{\mu\nu}$ and $R_{\mu\nu\l\k}R^{\mu\nu\l\k}$ are finite. In other words, both the $g_s$ and the
$\alpha'$ expansions are under control and the background 
is trustable in the far UV. Notice that there is a one-cycle, labelled by
the coordinate $\psi$ in $\omega_3$, that shrinks to zero size in the
large-$\r$ regime. This implies that strings wrapping this cycle will become
light and will enter the spectrum of the dual QFT at high energies. 

The dual dilaton is defined as $e^{2\hat\Phi}=\frac{e^{2\Phi}}{\mathcal{W}}$ where
\beq
\mathcal{W} = 3c_+ N_c\sqrt{12\rho-\frac{3}{2}}e^{8\rho/3}
\eeq
asymptotically, and so the dilaton is UV vanishing.

\subsection{IR asymptotic behaviour}
Let us now study the small radius regime of the metric, corresponding with
the low energy regime of the dual QFT. Things are a bit less-simple.
At leading order, terms in the metric depend explicitly of the 
original IR-parameters of the Baryonic Branch solution, but they also depend
on the values of the $v_2,v_3$ coordinates. The dual vielbeins in the IR tend to
\beq
\begin{array}{l l}
e^{\hat{1}}&=-\frac{32e^{\Phi_0/2}\sqrt{\mathcal{F}}h_1^{3/2}}{e^{2\Phi_0}\mathcal{F}^2+128h_1^2(v_2^2+v_3^2)}\bigg(v_3(dv_2+v_2\o_3)+v_2(v_2\o_2-\frac{1}{2\sqrt{2}}dv_3)-v_3^2(\o_1-\o_2)\bigg)\\
e^{\hat{2}}&=-\frac{2e^{\Phi_0/2}\sqrt{\mathcal{F}}\sqrt{h_1}}{e^{2\Phi_0}\mathcal{F}^2+128h_1^2(v_2^2+v_3^2)}\bigg(\sqrt{2}v_3\mathcal{F}e^{3\Phi_0}\o_1-\sqrt{2}v_2\mathcal{F}e^{\Phi_0}\o_3+\\
&~~~~~~~~~~~~~~~~~~~~~~~~~~~~~~16h_1\big(v_3dv_2-v_2dv_3 +(v_2^2+v_3^2)\o_2\big)\bigg)\\
e^{\hat{3}}&=-\frac{2e^{-\Phi_0/2}\sqrt{\frac{h_1}{\mathcal{F}}}}{e^{2\Phi_0}\mathcal{F}^2+128h_1^2(v_2^2+v_3^2)}\bigg(\sqrt{2}\mathcal{F}^2e^{2\Phi_0}(\frac{1}{2\sqrt{2}}dv_3-v_2\o_2)-16h_1v_2\mathcal{F}(v_2\o_3-v_3\o_1)+\\
&~~~~~~~~~~~~~~~~~~~~~~~~~~~~~~\sqrt{2}128 h_1^2v_3(v_2dv_2+v_3dv_3)\bigg)
\end{array}
\eeq
where we have defined
\beq
\mathcal{F}^2=4 (2)^{3/2}(h_1^{5/2}-2\sqrt{2}e^{\Phi_0}h_1)
\eeq
for convenience. The function $\mathcal{W}$ tends to
\beq
\frac{\mathcal{F}e^{\Phi_0}}{512 h_1}\bigg(\mathcal{F}^2e^{2\Phi_0}+128(v_2^2+v_3^2)\bigg)
\eeq

Here again, it happens that the dilaton is bounded and the Ricci scalar
and Ricci and Riemann tensors squared are finite.
This was expected, as we are performing a duality transformation on a space
that in the small-$\rho$
regime was of finite size (the $S^3$ in the deformed conifold). 
Dualities typically
invert `sizes' (or couplings). This example is not an exception.
One may start with a background solution where Supergravity is a good
approximation and  obtain that in the far IR
the new generated solution is still a trustable Supergravity background.

A point that we want to emphasize again is that in the far IR, the parameter
that was labeling the different `positions' on the Baryonic Branch
(that is the different baryonic VEVs) still 
appears in the  small-radius expansion above.
There is a still a one-parameter family of solutions. 
Indeed, notice the  dependence on the integration constants
$e^{\Phi(0)}$ and $h_1$ as defined in \cite{Gaillard:2010qg}, both
related to the number parametrising the Baryonic Branch.

\section{$SU(2)$ Structure of the background}\label{G-structures}
We will now study the associated G-structure with this solution. Again,
we will postpone details to the Appendix \ref{detailsBB}.  The geometry supports two pure spinors given by 
\beq
	\begin{aligned}
		\Phi_+ &= \frac{e^{A}}{8} e^{i \theta_+} e^{-i v\wedge w} \big( k_{\|} e^{-ij} - i k_{\perp} \omega \big) \ , \\
		\Phi_- &= \frac{i e^{A}}{8} e^{i\theta_-} (v+i w)\wedge \big( k_{\perp} e^{-ij} + i k_{\|} \omega \big) \ . 
	\end{aligned}  
\eeq
In the case at hand we find  
\beq
        \begin{aligned}
                &e^{2A} = \frac{e^{\Phi}}{\mathcal{C}} \\
                &\theta_+ = 0,~~~ \theta_-= \zeta(r) \\
                &k_{\|} = \frac{\sin \alpha}{\sqrt{1+\zeta.\zeta}} 
\qquad k_{\perp} = 
\sqrt{\frac{\cos^2 \alpha + \zeta.\zeta}{1+\zeta.\zeta}} \\
                &z = w - i \, v = \frac{1}{\sqrt{\cos^2 \alpha + \zeta.\zeta}} 
\big( \sqrt{\Delta} \tilde e^3 + \zeta_2 \sin\alpha 
\tilde e^\theta + i (\sqrt{\Delta} \tilde e^\r 
+ \zeta_2 \sin \alpha \tilde e^{\varphi}) \big) \\
                &j = \tilde e^{\r3} + \tilde e^{\varphi \theta} 
+ \tilde e^{21} - 
v \wedge w \\
                &\omega =  \frac{i}{\sqrt{\cos^2 \alpha + \zeta.\zeta}} 
\big( \sqrt{\Delta} 
(\tilde e^\varphi+ i \tilde e^\theta) - \zeta_2 \sin\alpha (\tilde e^\r 
+ i \tilde e^3) \big) \wedge (\tilde e^2 + i \tilde e^1).
        \end{aligned}
\label{su2structurezz}\eeq

Here the frames $\tilde{e}$  are obtained by a rotation, given by \eqref{eq:RmatixD5},  of those in \eqref{eq :BBDualVeil} and the parameters $\Delta, \zeta_i$ which enter into this rotation are specified by \eqref{eq: zetaBBsual}.

There are various immediate things to observe.
If we move to the large radius region of the geometry, the functions
$\sin\alpha(\r)\sim a(\r)\sim b(\r)\to 0$. The formulas simplify
and we obtain, among other things that $k_{\|}\to 0$. This implies that, as  happens in the paper \cite{Barranco:2013fza},
the two pure spinors are 
`perpendicular' in the large radius regime of the solution and 
the $SU(2)$-structure is {\it static}.  Similar behaviour was found in \cite{Macpherson:2013zba}, where a dynamical $SU(3)$-
structure in 7-d becomes orthogonal in the UV.
This changes as we evolve to the small radius regime of the background,
the $SU(2)$-structure is said to become {\it dynamical}.
In Section \ref{correspondenceqft}, we will discuss the physical effects
that are associated with a change in the $SU(2)$-structure, from {\it static}
 in the far UV to {\it dynamic} in the IR. 
%
 
\section{Correspondence with Field Theory}\label{correspondenceqft}
In this section, we will connect our previous geometrical studies with aspects of the quantum field theory that our background is dual to. As it was anticipated in  the paper
\cite{Itsios:2013wd}, we believe that the field theory dual to our massive IIA background should be a non-conformal version of the Sicilian gauge theories presented in 
 \cite{Benini:2009mz,Bah:2012dg} or the linear quiver field theories studied in  \cite{Aharony:2012tz}. There are certain things that can be inferred immediately, like for example the confining character of the QFT. This follows from the fact that the calculation of the Wilson loop will proceed exactly as in the case of the Baryonic Branch field theory. Indeed, the 
$R^{1,3}\times \rho$ part of the geometry is unchanged, hence, the Wilson loop will give the same result as before the non-Abelian T-duality. Nevertheless, many calculations done with the Klebanov-Strassler/Baryonic Branch background involved the `internal' five dimensional space. The purpose of this section will be to learn how some of those calculations for field theory observables change (or not) for the new geometries in massive IIA.

The idea that will guide us is that for a given correlation function or related QFT observable, that in the original background was calculated in a  way that is `independent' of the $SU(2)$ isometry used to perform the non-Abelian duality, will give the same result in the transformed background. We can think about those operators or correlators as `uncharged' under the $SU(2)$ symmetry in question. Ideas of this sort already worked in other solution generating techniques, like T-s-T dualities. Similar ideas also appeared in large $N_c$ (planar) equivalences between parent-daughter theories. The Physics of the common or `uncharged' sector goes through to the new field theory. The rest of the paper deals with observables that are, in principle `charged' under the $SU(2)$ symmetry.

In the paper \cite{Itsios:2013wd} it was shown that 
the cascade of Seiberg dualities--defined geometrically as 
a large gauge transformation of 
the NS two form and its effect on Page charges, 
persisted in the massive IIA background. 
In the paper \cite{Barranco:2013fza}, we started to
geometrise some of the field theory effects 
corresponding to the Klebanov-Witten non-Abelian T-dual. 
In the rest of this section, we will 
focus our attention on the relation between 
the dynamical character of the $SU(2)$-structure and 
the field theoretical phenomena of 
confinement and discrete R-symmetry breaking. 
We will show how the presence of Domain Walls 
with an induced Chern-Simons dynamics on their 
world-volume follows as a consequence of the confinement and the
dynamical character of the $SU(2)$-structure. 
Then, we will make clear  that the symmetry associated with changes in the $\psi$-direction is related with an anomalous $U(1)_R$ R-symmetry in the field theory. We will
define an instantonic object using an euclidean D0 brane; this will lead us to a  possible definition for a $\Theta$-angle and gauge coupling. We will find that this coupling has a non-conventional running in the far UV. We will then move into studying different aspects of the `baryonic branch', also present in our new backgrounds. We will find that a given fluctuation of the RR  background fields can be put in correspondence with a global continuous symmetry that the IR dynamics breaks spontaneously. We will find the associated Goldstone boson and an expression for the conformal dimension of such a baryonic operator.

\subsection{Dynamic SU(2): A pathway to confinement}
 In this section, we will make more concrete the relation between the QFT phenomena of confinement and the dynamical character of the $SU(2)$-structure. The first observation is that the `parallel projection' between both spinors, represented by $k_\|$ in eq.(\ref{su2structurezz}),
 is proportional to the quantity $\sin\alpha$. This quantity is related to the background functions as can be read from Appendix B of the paper \cite{Casero:2006pt},
\beq
\sin\alpha(\rho) = \frac{4 a e^{h-g}}{\sqrt{a^2+ 2 a^2 (4 e^{2h-2g}) + (4e^{2h-2g}+1)^2}}.
\eeq
This is compatible with the expression in eq.(\ref{cosinh}) after
following the algebra in Appendix B of the paper \cite{Casero:2006pt}.

The presence of the functions $a(\r), b(\r)$ in the Baryonic branch solution--see eqs.(\ref{vielbeinafter})-(\ref{eq: fidef})---are responsible for the de-singularisation of the
space (the appearance of a finite size $S^3$) and the IR minimization of the dilaton and warp factor. These have as a consequence the linear law, $E_{QQ}=\sigma L_{QQ}$ for large distance separations between the 
quark-antiquark pair. In other words,  the functions $a(\r), b(\r)$ and their effects on the warp factor and dilaton  
'produce' confinement. In the same vein, at the level of the metric, the presence of $a(\r)$ implies the breaking of the symmetry $\psi\to \psi+\epsilon$ into $\psi\to \psi+2\pi$. This is the remaining $\mathbb{Z}_2$ symmetry after the spontaneous discrete 
R-symmetry breaking. So, we see clearly that confinement and spontaneous R-symmetry breaking go hand-in-hand with the function $a(\rho)$. Hence, these  phenomena in the dual QFT are closely related to the presence of $k_\|$, which as we made clear is related to the dynamical character of the $SU(2)$-structure.
In the papers \cite{Apreda:2001qb,Mueck:2003zf}, the point was made  that the functions $a(\r), b(\r)$ were directly related with the gaugino condensate. This  suggests that in our massive IIA picture, there exists a relation of the form $<\lambda\lambda>\sim k_\|$. Similar ideas will be discussed in the paper \cite{NCE}.

\subsection{A comment on domain walls}
It was proposed in \cite{Itsios:2013wd}, that domain wall 
objects were realised in the Non-Abelian T-dual of the geometries
we are considering, as D2 branes that extend on $R^{1,2}$.
Indeed, the  induced metric, action and  tension of a 
$(2+1)$-dimensional object are,
\bea
& & ds_{ind}^2= e^{\Phi}\hat{h}^{-1/2}(- dt^2 + dx_1^2+dx_2^2),\nonumber\\
& & S_{BI}= -T_{D2}\int d^3 x e^{\Phi/2}\hat{h}^{-3/4},\;\;\; T_{DW}= 
T_{D2} e^{\Phi/2}\hat{h}^{-3/4}|_{\rho=0}.\nonumber
\label{inducedactiontension}\eea
If we also turn on a gauge field in the world-volume of this D2 
brane, a Chern-Simons-Maxwell action will be induced, at leading
order in $\alpha'$ on this D-brane,
\beq
S_{BIWZ}=-T_{D2}\int d^{2+1}x e^{\Phi/2}\hat{h}^{-3/4}
\sqrt{1-\alpha' F_{\mu\nu}F^{\mu\nu} } + T_{D2}\int d^{2+1}x F_0 A_1 \wedge F_2.
\eeq
We have used that a new WZ-like term appears in Massive IIA as explained in
\cite{Green:1996bh}. The Chern-Simons term is quantised, 
being proportional to $T_{D2}N_c$.\footnote{Note that it is the presence of an $F_0$ that allows D2 branes to be interpreted in this way, by way of comparison in \cite{Acharya:2001dz} the relevant branes with Chern-Simons dynamics are D4 branes with a bulk $F_2$ turned on. } 

In the type IIB Baryonic Branch solution(s), domain walls 
were realised by D5-branes extended on $R^{1,2}$ and the three-sphere
$\tilde{S}^3= [\tilde{\theta},\tilde{\varphi},\psi]$. Once a gauge 
field is turned on, a Chern-simons terms was induced, proportional to
$T_{D5}\int_{\tilde{S}^3} F_3$. Naively, we can think that both objects
are 'connected' by the non-abelian T-duality, under which the directions
on $\tilde{S}^3$ disappear and 
we are left with a D2 brane as described above.

Supersymmetry gives support to this. Indeed, 
around eq.(6.19) of the paper 
\cite{Martucci:2005ht}, we are presented with 
the calibration form for a domain-wall like object, which is given by the
real part of the pure spinor $\Psi_+$. Using that 
$|a|^2= e^{A}=e^{\Phi/2}\hat{h}^{-1/4}$, we obtain that
the BI action equals the calibration form. Notice also that
this selects the $k_\|$ component of the pure spinor.

As it was shown in the paper
\cite{Itsios:2013wd}, once the R-symmetry is broken in the Type IIB set-up, 
the non-abelian T-duality
maps these backgrounds to their partners in Massive IIA. In 
a minimally SUSY quantum field theory, the presence 
of domain-walls is tied up with confinement and the spontaneous breaking
of the $\mathbb{Z}_{2N_c}$-symmetry. As we emphasized,  these phenomena are related to
the `dynamical' character of the $SU(2)$-structure, hence to the presence of
the $k_\|$ part of the pure spinor.

\subsection{The fate of the $U(1)_R$ anomaly}
In the backgrounds  presented in \cite{Itsios:2013wd} and those of this paper it is somewhat natural to expect that the coordinate $\psi$ is singled out as being related to an R-symmetry of any putative field theory dual.    That this is true is by no means obvious, 
after all in the technical process of dualisation the fact that we retained the coordinate $\psi$ was  purely a result of a judicious gauge choice.   Here we provide evidence that this is indeed the correct identification and furthermore that this $U(1)$ is afflicted with an anomaly, 
breaking it down to a discrete subgroup.  

A robust understanding of how $\partial_\psi$ plays the role of the  R-symmetry in the holographic dual was given in   \cite{Klebanov:2002gr}  with several important details of the supergravity solution clarified in \cite{Krasnitz:2002ct}.  The essential point of \cite{Klebanov:2002gr} is to introduce a bulk 5d gauge field  that gauges this  $U(1)_\psi$ by  making the replacement $d\psi \rightarrow \chi = d\psi - 2 A$ in the metric.  This must be supplemented with an appropriate ansatz for the fluxes. In the case of the Klebanov-Witten background one finds that the resultant gauge field is massless and is the dual fluctuation to the global $U(1)_R$ of the gauge theory.   However, in the non-conformal cases, the correct ansatz for the fluxes actually yields a massive gauge field (the mass here comes from a St\"uckelberg rather than Brout-Englert-Higgs mechanism).   
 
Let us begin our discussion with the non-abelian T-dual of the Klebanov-Witten backgound. The NS sector of the geometry is given by  
\begin{equation}
\begin{aligned}
d  s^2 &=   ds^2_{AdS_5} + \frac{1}{6}ds^2_{S^2}  + \frac{6 v_2^2}{\Delta}   \s_{\hat 3}^2+ \frac{6}{\Delta} \left[ (1+27 v_2^2) dv_2^2 + 54 v_2 v_3 dv_2 dv_3 + \frac{3}{4}\left(\Delta -54 v_2^2 \right)dv_3^2\right]  \ ,
  \nonumber  \\
   B_2 &= \frac{18 \sqrt{2}}{\Delta} v_2 v_3  \s_{\hat 3}\wedge dv_2 + \frac{\left(\Delta -54 v_2^2 \right)}{\sqrt{2} \Delta}  \s_{\hat 3}\wedge dv_3  \ , \\ 
 e^{2 \Phi} &= 81 \Delta^{-1}  = 81 \left( 2+ 54v_2^2 + 36v_3^2  \right)^{-1} \ ,
 \label{T11dual}
\nonumber
\end{aligned}
\end{equation} 
where $ \s_{\hat 3} = d\psi + \cos\th  d\phi $.  This  metric is supported by   RR two and four form fluxes.   The $U(1)$ acting as $\partial_\psi$ can be gauged by making the replacement $ \s_{\hat 3} \rightarrow \tilde{\chi}=\s_{\hat 3}  - 2 A$  in the  NS sector above.  The potentials corresponding to the correct modification of the RR forms that support this fluctuation are  given by 
\begin{equation}
\begin{aligned}
 C_1 &= -\frac{2 \sqrt{2}}{27}  \left( \cos\theta d\phi + A \right) \ ,  \\
 C_3 &=-\frac{2}{27} v_3 \tilde\chi\wedge(\tilde{\omega}_2 - dA ) +\frac{2}{9} v_3 \star_5 dA \ ,
 \end{aligned}
 \end{equation}
 where we introduce the volume form on the $S^2$, $\tilde{\omega}_2 = \sin\theta d\theta d\phi$ and $\star_5$ is the Hodge dual in the $AdS_5$ directions.    This solves the linearised equations of motions, linearised Einstein equations and Bianchi identities provided that the gauge field obeys the equation $d\star_5 dA$.    This, together with the fact that the Killing spinors of the geometry are charged under $U(1)_\psi$ identifies this as the dual to the R-symmetry.   Upon substitution of this ansatz in to the action   one finds all the gauge field dependance gives a field strength squared  contribution,    
\begin{equation}
\delta S = f(v_2 ,v_3)  F_{\mu \nu} F^{\mu \nu}  
\end{equation}
for some function $f(v_2,v_3)$  of the internal coordinates that will be integrated over in a reduction to a five-dimensional theory. 

Now we turn to the non-conformal geometry obtained by transformation of the Klebanov-Tseytlin geometry (since we are only interested in the UV behavior we will not need the full Klebanov-Strassler or baryonic branch).  The NS sector, with the $U(1)_\psi$ gauged, is given by 
   \def\cV{{\cal V}_3} 
   \begin{equation}
 \begin{aligned}\label{dualKT}
 ds^2 &= h^{\frac{1}{2}} dr^2
+  h^{-\frac{1}{2}} ds^2_{R^{1,3}} + \frac{ r^2 h^{\frac{1}{2}}}{6}ds^2_{S^2}  + \frac{6 r^4 h v_2^2}{\Delta}\tilde{\chi}^2\\
&\qquad  + \frac{6}{\Delta} \left[ (r^4 h +27 v_2^2) dv_2^2 + 54 v_2 \cV dv_2 dv_3 + \frac{3}{4}\left(\frac{\Delta}{r^2 h^\half} -54 v_2^2 \right)dv_3^2\right] \\ 
B_2 & = \frac{18 \sqrt{2}}{\Delta} v_2 \cV \tilde\chi \wedge dv_2 + \frac{\left(\Delta -54 r^2 h^\half v_2^2 \right)}{\sqrt{2} \Delta} \tilde\chi\wedge dv_3 +\frac{r^5 h'(r)}{54 M }\tilde{\omega}_2 \\
 e^{2 \Phi} &= 81 \Delta^{-1}  = 81 \left( 2r^4 h + 54v_2^2 + 36\cV^2  \right)^{-1} \ .  
    \end{aligned}
  \end{equation} 
Here $h(r)$ is the usual Klebanov-Tseytlin warp factor and  $\cV= v_3+\frac{ r^5 h'(r)}{27 \sqrt 2 M}$.  Without the gauging this is a solution of massive IIA with Romans' mass proportional to $M$. By examining how the non-abelian T-duality transformation acts on the ansatz given by Krasnitz in  \cite{Krasnitz:2002ct}, we can determine a suitable ansatz for the fluxes: 
     \begin{equation}
 \begin{aligned}
   C_1 &= -\frac{M}{2}v_3 \cos\theta d\phi +\frac{M}{2}\psi dv_3  -2 \sqrt{2} K_1 -\sqrt{2} C_0 \left(\cV dv_3 + v_2 dv_2  \right) \\
 C_3 &= 2 \cV K_3 - \frac{ M \sqrt{2}}{4} \psi \tilde\omega_2 \wedge\left(v_2 dv_2 + v_3 dv_3 \right)  \\
 &  + \frac{ 2\sqrt{2}}{M} f(r)  C_0 \tilde\omega_2 \wedge\left(v_2 dv_2 + \cV  dv_3 \right)  -2 v_3 \tilde\chi \wedge dK_1   -4v_3 \tilde\omega_2 \wedge K_1  + \Theta_3
   \end{aligned}
  \end{equation} 
   The remaining term in the three-form potential is given implicitly by\footnote{The exterior derivative of right hand side of this expression vanishes on the   equations \eqref{K3eqs}.} 
 
   \begin{equation}
 \begin{aligned}
 d\Theta_3 = \frac{1}{\sqrt{2}} M  h^\frac{1}{4} \star_5 \left( C_0  dr + \frac{2}{3} r W  \right) + \frac{3 M}{\sqrt{2}} dr \wedge K_3 \ . 
     \end{aligned}
  \end{equation}
 Here $W$ is a gauge invariant 1-form that combines the gauge field $A$ with a St\"uckelberg scalar scalar $W = A - d\lambda$ though for practical purposes we follow \cite{Krasnitz:2002ct} and  chose a gauge in which $W=A$.    This is a solution to the linearised flux equations and Bianchi identities provided the fields introduced obey the constraints on the ansatz required in \cite{Krasnitz:2002ct}:
   \begin{equation}
 \begin{aligned}\label{K3eqs}
 K_3 &= - \frac{3}{r h^\frac{1}{4} }\star_5 dK_1 \ ,  \\ 
 dK_3 &= \frac{24} {r^3 h^\frac{3}{4}} \star_5 \left( K_1 + f(r) W \right) \ , \\
 0 &= \frac{1}{3} \partial_r \left( r h^{-1}  W_r \right) + \frac{r}{3} \partial_i W_i + \frac{1}{2} \partial_r (h^{-1} C_0) -\frac{36}
 {r^4h^2} \left( (K_1)_r + f(r) W_r \right) \ , \\ 
 0 & = \frac{1}{54} \partial_r \left( r^5  \partial_r C_0 \right) + \frac{r^5 h}{54} \partial_i \partial_i C_0 - \frac{M^2}{2 h} W_r - \frac{3 M^2}{4 h r } C_0 \ . 
    \end{aligned}
 \end{equation}
Here $\star_5$ is the Hodge dual with respect to the metric $ds_5^2  = h^{\frac{1}{2}} dr^2+  h^{-\frac{1}{2}} ds^2_{R^{1,3}}$.  In  \cite{Krasnitz:2002ct} it was shown how these equations \eqref{K3eqs} can be diagonalised   by defining 
 \begin{equation}
 W^1 = W - \frac{54}{hr^4} K_1 \ , \quad W^2 = W + \frac{27}{h r^4} K_1 \ . 
 \end{equation} 
The mode $W^1$ corresponds to a massive gauge field whose mass as a result of the spontaneous (anomalous) breaking of R-symmetry.  The mass of this mode is given by  \cite{Krasnitz:2002ct}:
  \begin{equation}
  m^2 = \frac{4}{\a'(3\pi)^\frac{3}{2} } \frac{(g_s M)^2}{ (\lambda N)^\frac{3}{2} } 
  \end{equation} 
The interpretation is identical here and we conclude therefor  that the $U(1)_R$ symmetry is anomalously broken.    
 
 \addtocontents{toc}{\protect\setcounter{tocdepth}{2}}  
\subsubsection{Dependance on $\psi$ in the potentials and D0 brane instantons}

To understand this breaking as an anomaly it is informative to look at the forms of the RR potentials.  For the non-Abelian T-dual of the Klebanov-Witten we have following potentials 
\begin{equation}
 \begin{aligned}
 C_1&=  \frac{N  \pi }{\sqrt{2}} \cos \theta d\phi \ , \\
 C_3 &= -\frac{N  \pi v_3 }{2} \sin \theta d\theta \wedge d\phi \wedge d\psi   \ . 
     \end{aligned}
\end{equation}
For the dual of the Klebanov-Tseytlin (which has Romans mass proportional to $M$) we have 
\begin{equation}
 \begin{aligned}
 C_1&= \frac{M}{2} v_3 \cos\theta d\phi - \frac{M}{2} \psi dv_3 \ ,  \\
 C_3 &= - \frac{\sqrt{2} M}{8 } \left(v_2^2 + v_3^2 \right)\sin \theta d\theta \wedge d\phi \wedge d\psi  \ . 
    \end{aligned}
\end{equation}
Note how the dependence on $\psi$ in $C_1$ is quite different in the potentials in the conformal and non-conformal cases. 

Let us now   consider D0 branes. These D0 branes will move in the $v_3$ direction, leaving all other coordinates
fixed, in particular we will choose $v_2=0$. We can then calculate using \eqref{dualKT}
the induced metric for this D0 brane, relevant gauge potential and its BIWZ action, that will read
\bea
& &  ds_{ind}^2=g_{v_3 v_3}dv_3^2= \frac{9}{2 r^2 h^{1/2}}dv_3^2, \;\;\; C_1= -\frac{M}{2}\psi dv_3,\nonumber\\
& & S_{BIWZ}=-T_{D0}\int dv_3 e^{-\Phi}\sqrt{g_{v_3 v_3}} + T_{D0}\int C_1=T_{D0}\int dv_3 \sqrt{\frac{r^2 h^{1/2}}{9}
 +\frac{2{\cal V}_3^2}{r^2 h^{1/2}}} - T_{D0}\frac{M\psi}{2}\int dv_3.\nonumber
\eea
We use now that $T_{D0}=\frac{1}{g_s \sqrt{\alpha'}}$. Also, we call $\sqrt{\alpha'}
L_{v_3}
= \int dv_3$, the dimensionless length of the $v_3$ direction.

We will equate the BIWZ action of this euclidean D0 brane with the gauge coupling and the $\Theta$ angle
imposing that $S_{BIWZ}=\frac{8\pi^2}{g^2}+ i\Theta$. In other words, we consider this D0 brane to be 
an instanton in the dual gauge theory.

Analysing the WZ term, we have that (like above, we choose $g_s=1$),
\beq
S_{WZ}= \frac{M}{2}\psi L_{v_3}= \Theta.
\eeq
Using that the theta angle
should be periodic, we can impose that the allowed changes in the angle $\psi$ get selected to be
\beq
\frac{M}{2}(\psi+\Delta \psi) L_{v_3}= \Theta +2 k \pi
\eeq
which implies that
\beq
\Delta \psi=\frac{4 k \pi}{M L_{v_3}} \ . 
\eeq
So, we see that there is a breaking of the global continuous symmetry into a discrete one.   The residual discrete symmetry is determined by the domain of the coordinate $v_3$.   {\it In the case in which we would like to impose this discrete symmetry to be the same as before the non-Abelian duality we should impose that $L_{v_3}=2$.}   Indeed, one of the major challenges with understanding non-abelian T-duality is to identify the periodicities of the coordinates of the T-dual geometry.  Here we see a direct link between a field theory property (the anomaly) and the global properties of the geometry.

Let us look at the BI term. We have that the gauge coupling, associated is
\beq
\frac{8\pi^2}{g^2}= T_{D0}\int dv_3\Big[ r^2 h^{1/2} +\frac{2}{r^2 h^{1/2}}(v_3 
+\frac{r^5 h'}{27\sqrt{2}M})^2  \Big]^{1/2} .
\eeq
We can perform the integral explicitly, but it is perhaps more illuminating to 
look at the large radius limit of the expression above. After all, we are doing this
calculation in the non-Abelian dual of the Klebanov-Tseytlin solution, 
we should only trust the result in the far UV. We have then, considering the leading
term in the large-$r$ expansion,
\beq
\frac{1}{g^2}\sim (\log r)^{3/2}
\eeq
this reproduces a result obtained by other means in \cite{Itsios:2013wd}.

\subsection{The fate of $U(1)_B$} 
The Klebanov-Witten 
$SU(N) \times SU(N)$ conformal field theory coming from D3 branes at the tip of the conifold has a $U(1)$ baryonic number symmetry acting  as $A_i \rightarrow e^{i \alpha} A_i$, $B_j \rightarrow e^{-i\alpha} B_j$.  In the gravity dual this number current gives rise to a massless $AdS_5$ gauge field
\begin{equation}
\delta C_4 = \omega_3 \wedge {\cal A} \ ,
\end{equation}
where $\omega_3$ is the usual closed three form on $T^{1,1}$.   The non-abelian T-dual of the $AdS_5\times T^{1,1}$  geometry was obtained in \cite{Itsios:2012zv}.  In the T-dual geometry, this $U(1)_B$ mode 
translates into 
a perturbation, which solves the linearised supergravity equations of motion, given by
\begin{equation}\begin{aligned}
& \delta C_1 = \frac{1}{9} {\cal A} \ ,  \\
& \delta C_3 =   W_2 \wedge  {\cal A}+ \frac{1}{9}  u du \wedge{\cal F}  + \frac{\sqrt{2}}{6} u dv_3 \wedge \star_4 {\cal F}  \ .
\end{aligned}
\end{equation}
The final two terms in $\delta C_3$ come from the a contribution from $\delta C_6$   under the T-duality transformation\footnote{For the  $AdS_5\times T^{1,1}$ we use $ds^2_{AdS}= du^2 + e^{2u} (\eta_{ij} dx^i dx^j)$.}.  Although the two-form  $W_2$ has a simple form
\begin{equation}
W_2 = \frac{v_3} {9} d\sigma_3  + \frac{\sqrt{2} v_2 e^{2\hat{\phi} }}{81} \sigma_3 \wedge(2 v_3 dv_2 - 3 v_2 dv_3   ) 
 \end{equation}
 it can not easily be written in terms of the invariant tensors that define the $SU(2)$ structure of the geometry.

 The existence of this mode is suggestive that the field theory duals corresponding to the conformal geometries constructed in \cite{Itsios:2012zv} have a global $U(1)$ symmetry in addition to the preserved $U(1)_R$.       In fact,  the geometry T-dual to the Klebanov Witten is closely related to those proposed in \cite{Bah:2012dg} as the gravity duals to ${\cal N}=1$ SCFT's formed by wrapping M5 branes on Riemann surfaces (which in this case is genus zero giving rise to many subtleties).  These SCFT's do indeed have $U(1)_R \times U(1)_F$ Abelian global symmetries which are  seen geometrically as isometries of the corresponding eleven-dimensional supergravity solution. Upon reduction to ten-dimension one of these U(1)'s gets degeometrized  corresponding to the above    gauge field $\delta C_1 = {\cal A}$. 
 
In this paper our main focus has been the cascading field theory where at the last  step of the cascade when the gauge group is $SU(M)\times SU(2M)$ the baryons acquire expectation values,
\beq
{\cal B}=i \xi \Lambda^{2M},\;\;\; \tilde{\cal{B}}= \frac{i}{\xi}\Lambda^{2M} \ .
\eeq
On this Baryonic branch the $U(1)_B$ symmetry is spontaneously broken.  To see this from the gravity perspective it  is sufficient to work with the Klebanov-Strassler geometry corresponding to the field theory at the $\mathbb{Z}_2$ symmetric point of the Baryonic branch.  As shown in \cite{Gubser:2004qj}, there is a massless glue ball corresponding to a 
Goldstone mode associated with changing the phase of $\xi$ which is given by\footnote{Here and elsewhere use the standard notation for the deformed conifold and Klebanov Strassler geometry which can be found e.g. in appendix of  \cite{Gubser:2004qj}.  For the KS we stick with the notation $\tau$ as the radial coordinate but will use  $r$ elsewhere.} 
\beq\label{eq:glueball}
\begin{aligned}
& \delta H = 0 \ ,  \\ 
& \delta F_3 = f_1 \star_4 da - d(f_2(\tau) da \wedge g^5 ) \ , \\
& \delta F_5 =f_1\left(  \star_4 da  -\frac{\epsilon^\frac{4}{3}}{6 K^2(\tau)} h(\tau) da \wedge d\tau \wedge g^5 \right) \wedge B_2 \ . 
\end{aligned} 
\eeq 
The linearised supergravity equations are solved when the pseudo-scalar is  a harmonic function in $\mathbb{R}^{3,1}$  and the  function $f_2(\tau)$ obeys a second order differential equation admitting a normalisable solution.

The non-abelian T-dual geometries considered also admits a similar mode, which can be obtained simply by performing a T-dualisation of the ansatz for the scalar modes in the seed IIB solutions.  The T-dual of the Klebanov-Strassler geometry   was  obtained explicitly in \cite{Itsios:2013wd}.  Performing a dualisation of the ansatz   \eqref{eq:glueball} gives rise to a perturbation $\delta F_2$ and $\delta F_4$.   This perturbation solves the supergravity equations of motion when $f_2$ obeys the same differential equation as for the ansatz \eqref{eq:glueball}.   The expressions for $F_2$ and $F_4$ are not particularly enlightening though for completeness let us   provide a few details.  Here we display the results in the   UV regime where the geometry is given by  \eqref{dualKT}.  The corresponding deformations to the potentials are given by
\begin{equation}
\begin{aligned}
& \delta C_1 = (2 v_3 f_2(r) + f_3(r)) da  \\
& \delta C_3 =  \left[ f_4(r) - \frac{f_1}{\sqrt{2}} \left(v_2^2 + (v_3 - \frac{N  \pi }{\sqrt{2}  M})^2   \right)\right] \star_4 da \\
& \qquad  \qquad \qquad - \frac{f_2}{\sqrt{2}} da  \wedge \sigma_3 \wedge d(v_2^2  + v_3^2)   - \frac{f_3}{ \sqrt{2}} da  \wedge \sigma_3 \wedge dv_3  + da \wedge\sin \theta d\theta \wedge d\phi \left(f_5 - \frac{v_3}{\sqrt{2}} f_3 \right)
\end{aligned} 
\end{equation}
The extra functions introduced above are completely determined by $f_1$ and $f_2$ according to 
\begin{equation}
\begin{aligned}
& f_1' = 0 \ , \quad  2 r^4 f_2'' = - 6 r^3 f_2' + 16 r^2 f_2 + 27 M^2 f_1 \log r /r_0  \ ,  \\ 
& f_3' =  \frac{1}{6} \left( -3 \sqrt{2} r f_1 h(r) \log r /r_0 - 2 T(r) f_2'   \right) \ , \quad  f_4' = \frac{2 \sqrt{2}}{3} r f_2 \ ,  \\
& f_5' =  \frac{1}{108} \left(-2 \sqrt{2}r^5 f_1 h(r) = 18 M r f_1 h(r)  T(r) \log r /r_0  - 3 \sqrt{2} T(r)^2 f_2'  \right)\ ,
\end{aligned} 
\end{equation}
where $T(r) = \frac{9}{\sqrt{2}}M \log r/r_0$ and $h(r) = \frac{27}{32 r^4} \left(3M^2 + 8 N  \pi + 12 M^2 \log r/r_0 \right)$.

 The existence of this mode suggests   a   spontaneously broken global $U(1)$ in the field theories dual to the  geometries obtained in section \ref{NATDBB}.   In the conformal case, the unbroken $U(1)$ becomes geometrized upon lifting to M-theory whereas these non-conformal backgrounds are solutions of {\em massive} IIA and so can not be lifted. This further underlines the expectation that a $U(1)$ is broken. 

In the same multiplet as the pseudo-scalar goldstone is a scalar perturbation corresponding to changing the magnitude of $\xi$.  In the same vein as above, one could deduce the  fate of this scalar perturbation under the T-duality transformation; it will give a similar, albeit complicated, perturbation in the dual IIA background.   Since the full baryonic branch geometry found  in \cite{Butti:2004pk} can be thought of as exponentiating such transformations to give arbitrary values of  the Baryonic vev,  implicitly in the geometries presented in section \ref{NATDBB}   we have already done just that.

\subsection{The fate of the baryon   condensate}

In Klebanov-Witten theory the closest analogy to a  baryon vertex - the object to which N {\it external}  quarks can attach  \cite{Witten:1998xy} - would be a D5 brane wrapping the $T^{1,1}$ space with world volume coordinates  $\{ x_0, \theta_1 ,  \phi_1 , \theta_2 , \phi_2 , \psi\}$ \cite{Benna:2006ib}.   The primary reason for this identification follows the argument made in  \cite{Witten:1998xy}; since we have 
\begin{equation}
\int_{T^{1,1}} F_5 \propto N \ , 
\end{equation}  
the  WZ term induces a charge to the world volume $U(1)$  gauge field ${\cal A}$     via the coupling
\begin{equation}
\int_{\mathbb{R} \times T^{(1,1)} }  {\cal A} \wedge F_5 \ .
\end{equation}  
 This introduces N units of charge which must be canceled by some other source to give zero net charge in a closed universe.  This cancelation is achieved by N elementary strings stretching from the boundary to the brane whose end points are external quarks.   A perhaps naive approach would be to suggest in the IIA geometry dual to the  Klebanov-Witten theory a similar role could be played by a D2 brane wrapping the $S^2$ with world volume coordinates  $\{ x_0, \theta , \phi \}$.  Indeed, since in the case of T-dual to Klebanov-Witten we have $C_1 \propto \cos \theta d\phi$ the WZ coupling ${\cal F} \wedge C_1$ produces a charge contribution  for the gauge field that could be cancelled with external quarks just as in the Klebanov-Witten scenario.  It would be of some interest to study the baryon vertex in the massive IIA backgrounds.\footnote{Before duality in the cascading theories this is a D3 brane and it seems quite possible that D0 branes might play this role of the baryon vertex in the cascading massive IIA geometries.  We thank O. Aharony and J. Sonnenschein for this suggestion.}
  
   This  baryon vertex should however  be distinguished from the configuration representing the actual baryon condensate - which should be supersymmetric, gauge invariant and not require BIon spikes.  The configuration that describes the baryon condensate is a Euclidean D5 brane wrapping the  $T^{1,1}$ and the radial directions \cite{Benna:2006ib}.  This D5 has D3 branes dissolved within \cite{Aharony:2000pp}  which are traded for a world volume gauge field.  Following the logic applied to the baryon vertex  one might   anticipate that in the IIA geometries presented here, the role of the condensate is played by  a wrapped Euclidean D2 brane on the $S^2\times \mathbb{R}$   with a world volume gauge field. 
 
   To determine the existence of such a configuration, rather than calculate the kappa symmetry projectors, we will harness the power of the G-structure and the calibration techniques of \cite{Martucci:2005ht}.  The condition for a supersymmetric Euclidean $p$ brane on a cycle $\Sigma$ is essentially the same as that of a Lorentzian $p+4$ brane that is spacetime filling in the Minkowski directions.   This condition is given by
\begin{equation}\label{calib}
e^{-\phi} \sqrt{-\det (g|_\Sigma+{\cal F} )} \, d^p\sigma  =  8 e^{3 A - \phi}  Im \Phi \wedge e^{-{\cal F}} |_\Sigma
\end{equation} 
where the world volume field strength is ${\cal F}= B|_\Sigma + 2 \pi \alpha' d {\cal A}$ and the pure spinor entering   the calibration form is   given  $\Phi = \Psi_+$ for IIB and $\Phi = \Psi_- $ for IIA.

Before looking at this question in the context of the full baryonic branch let us address it in the conformal case - we would still anticipate a supersymmetric configuration to exist even.  In the Klebanov-Witten theory the E5 configuration of 
a brane extended along 
$\Sigma = \{r, \theta_1, \phi_1 , \theta_2 , \phi_2 , \psi\}$   
with a world volume gauge field
\begin{equation}
{\cal A} = \frac{1}{3} \zeta(r)  \left(d\psi +   \cos \theta_1 d\phi_1 + \cos \theta_2 d\phi_2 \right) \ , 
\end{equation} 
 obeys the calibration condition \eqref{calib} provided that
 \begin{equation}\label{E5gauge}
 \zeta  \zeta' = \frac{1}{4}  - \zeta^2 \ , 
 \end{equation}
 which of course can be readily integrated. 
  
 In the IIA non-Abelian T-dual of
the Klebanov-Witten geometry we find 
an E2 configuration extended along 
$\Sigma = \{r, \theta , \phi\}$ at  the point $v_2=0$ 
but with a non-trivial embedding $v_3 = f(r)$.  
We search for  a supersymmetric configuration solving the calibration condition \eqref{calib} when supported by a gauge field 
 \begin{equation}
 {\cal A} = \frac{1}{  \sqrt{2}  } \alpha(r)  \cos\theta d\phi  \ . 
 \end{equation} 
 From the calibration condition one finds firstly 
that the embedding $f(r)$ and the gauge field should differ only by a constant $c_0$.  The gauge field should then obey an equation 
 \begin{equation}
 \alpha'(r) = \frac{1- 18 c_0 \alpha -18  \alpha^2}{9(c_0 +2 \alpha)} 
 \end{equation} 
 which can also be readily solved and one notices that when $c_0 =0$ has the same form as the equation \eqref{E5gauge} governing the configuration in IIB.

Lets move up to the KT geometry working in the exact logarithmic solution\footnote{This is considerably simpler than the deformed conifold of the KS and  reproduces all the main features of the calculation in   \cite{Benna:2006ib} with the conformal dimension of the condensate agreeing to leading order.  Using the calibration technique we checked that the resultant gauge field equation of motion  agrees exactly with that of \cite{Benna:2006ib}.  }.  First we recapitulate the calculation for the baryon condensate in the IIB background.   Using the calibration technique one readily finds the E5 configuration is the same but with the gauge field equation of motion eq. \eqref{E5gauge} modified to be 
\begin{equation}
\zeta'(r) = \frac{2 r^4 h(r) + T(r)^2 - 8 \zeta(r)^2}{8r \zeta(r) } \ ,  
\end{equation} 
where $T(r) = \frac{9}{\sqrt{2}}M \log r/r_0$ and $h(r) = \frac{27}{32 r^4} \left(3M^2 + 8 N \pi + 12 M^2 \log r/r_0 \right)$. This equation may be integrated to yield 
\begin{equation}
\zeta(r) = \frac{9M}{8r\sqrt{2}} \left( c + 3r^2- 4 r^2 \log(r) + 8 r^2 \log(r)^2   \right)^{\frac{1}{2}}  \ , 
\end{equation}
where $c$ is a constant of integration which we now set to zero since its contributions are in any case sub-leading.  Inserting this into to the DBI action one finds, changing variables to $t= \log r$, 
\begin{equation}
S_{E5} = \tau_5 vol(T^{1,1}) \int^{t_{UV}} dt \frac{27M^3}{64\sqrt{2}}  (1+2 t^2 +8t^3) (3- 4 t + 8 t^2)^{\frac{1}{2}}  \ . 
\end{equation} 
In \cite{Benna:2006ib},   $e^{- S_{E5}}$  was identified with the bulk field dual to the baryonic condensate. Using the standard asymptotic expansion the field theory scaling  dimension can be extracted (at least in the large $t$ regime) as  
\begin{equation}
\Delta(r)  =  \frac{d  S_{E5}} {d \log r}   = \frac{27}{16\pi^2} M^3(\log r)^2 + {\cal O}(\log r)  \ , 
\end{equation} 
reproducing exactly the  result of \cite{Benna:2006ib} notable for the scaling dimension dependence on the energy scale of the baryons as anticipated from the field theory.  

In the non-abelian T-dual the situation is already rather involved.  We search for an E2 configuration extended along $\Sigma = \{r, \theta , \phi\}$ at  the point $v_2=0$ and now with $ v_3 = \chi(r)$ and an ansatz for the gauge field 
 \begin{equation}
 {\cal A} = \frac{1}{  \sqrt{2}  } \alpha(r)  \cos\theta d\phi  \ . 
 \end{equation}
We take the  square of the calibration equation eq.~\eqref{calib}  and first consider terms proportional to $\cos^2\theta$. From these one finds a first equation relating the gauge field and the embedding in $v_3$: 
\begin{equation} 
\alpha'(r)  =  \chi'(r) \ . 
\end{equation} 
We let $c_0  $ be the additive constant between $\alpha$ and $\chi$.  Then from the remaining terms in  eq.~\eqref{calib}  one finds a differential equation for the gauge field 
\begin{equation}\label{eq:alpha}
r \alpha'(r)  = \frac{ 1}{18 (  c_0 +2 \alpha ) }  \left(2 r^4 h(r)  - 6 c_0 T + T^2 - 36 c_0  \alpha - 36 \alpha^2   \right)  \ . 
\end{equation}
Changing variable to $t = \log r $ one can solve this equation on the exact logarithmic solution:
\begin{equation}
\alpha(r) = -\frac{c_0}{2} \pm \frac{r^{-3/2} }{8}\left[64 r c  +  r^3\left(16 c_0^2 +  3M(8 \sqrt{2} c_0 + 9 M - 4(4\sqrt{2} c_0 +3M)\log r + 24 M \log r^2  )  \right)\right]^\half 
 \end{equation}
 here $c$ is an integration constant giving sub-leading contributions that we hence ignore. 
 
 Using the equation \eqref{eq:alpha} we find that the DBI action is given by 
  \begin{equation}
S_{DBI} = \kappa  \int  \frac{dr}{r}\frac{1}{648} (c_0 + 2 \alpha)^{-1}\left( 2 r^4 h + (T+ 6\alpha)^2 \right) \left( 2 r^4 h +  (T-  6(c_0+ \alpha))^2 \right) \ .
\end{equation}
If we expand out asymptotically we find that  
  \begin{equation}
S_{DBI} \sim \kappa  \int^{t_{UV}}  dt  \frac{27 M^3 t^2 }{8\sqrt{2}}  + \frac{9M^2 t }{32}  \left(3 \sqrt{2} M -4 c_0 + 8\sqrt{2}\frac{ N}{M} \pi \right) + {\cal O}(t^{0}) \ , 
\end{equation}
which suggests an operator with a scaling dimension 
\begin{equation}
\Delta =  \frac{27 \kappa M^3  }{8\sqrt{2}}    (\log r)^2
\end{equation} 
where $\kappa = T_{D2} \rm{vol}(S^2) = \frac{1}{\pi}$.   It would be interesting to pursue this line of reasoning further by 
extracting the value of the condensate across the baryonic branch.  This is technically rather involved and we do not intend to do so in this report. 
   
\section{Conclusions and Future Directions}\label{conclusions}
In this paper we have examined a new family of  solutions of massive IIA supergravity. 
These new backgrounds were obtained by performing a
non-abelian T-duality on the geometry that describes the non-perturbative Physics of
 the baryonic-branch of the Klebanov-Strassler field theory.  
We have explored the transition from $SU(3)$ structure, 
characterising the `seed' backgrounds to
the dynamical $SU(2)$-structure that describes 
the resulting massive IIA solutions.
 We made clear--at least for the type of backgrounds
studied here-- that the dynamical character of the $SU(2)$ structure
is directly related to the phenomena of confinement and symmetry breaking. 
We believe that all these new features have not been discussed
in previous literature, in a context as clear and 
unifying as the one presented here.

The new backgrounds discussed in this paper display
a host of interesting non-perturbative phenomena that `define' 
the dual field theory.
Some of these are,

\begin{itemize}
\item{The non-conformality of the geometry is enabled by a non-zero Romans' mass.}
\item{Whilst the UV geometries proposed in  \cite{Itsios:2013wd} are characterized by {\em static} $SU(2)$ structure \cite{Barranco:2013fza} the full IR complete geometry of this paper has {\em dynamic} $SU(2)$ structure.} 
\item{The transition to dynamic $SU(2)$ structure gives a geometric realization of confinement  and permits 
supersymmetric D2 branes that act as domain walls in the IR. This realises geometrically the relation between confinement,  the spontaneous breaking of a discrete R-symmetry  and the presence of
domain walls.}
\item{The $U(1)_R$ symmetry is realized by the vector $\partial_\psi$ and the corresponding fluctuation, which is a  massless gauge field in the conformal case, acquires a mass indicating an anomalous breaking. }
\item{Euclidean `instantonic' branes reproduce this anomaly of the R-symmetry and at the same time suggest a non-conventional running for a suitably defined gauge coupling.}  
\item{ A further $U(1)$ (baryonic) symmetry  is broken. In the conformal case of \cite{Itsios:2013wd} this symmetry is unbroken and is realized geometrically by the M-theory circle. In our backgrounds, once conformality is broken
by the addition of fractional branes, the symmetry is no longer geometrical as we are now in a massive IIA context. The 
$U(1)_B$ symmetry  is spontaneously broken and we identified a corresponding massless glueball (the associated Goldstone boson).} 
\item{ We give evidence that this $U(1)_B$ may be thought of as baryonic and that a baryonic condensate is given by a Euclidean D2 brane wrapping a two-cycle in the geometry.}
\end{itemize}
 Although we do not yet have a complete understanding of the field theory dual to this new geometry, the results of this paper together with those in \cite{Itsios:2013wd} suggest that it may be a non-conformal and cascading version of the Sicilian 
theories of  \cite{Benini:2009mz,Bah:2012dg} or the linear quivers of
 \cite{Aharony:2012tz}.  
 
 We would like to close this paper on a 
forward looking note.  We suggest that the  features mentioned 
above may be prototypical of a   
wider class of holographic duals.  
 The  theories in  \cite{Benini:2009mz,Bah:2012dg}
 and also the IIA linear quivers of \cite{Aharony:2012tz}, 
present a wide new class of interesting examples of ${\cal N}=1$ SCFTs.
 We anticipate that by a modification of these theories 
(this paper suggests that the modification will involve adding D8 branes in IIA) 
one can obtain a variety of  non-conformal gauge theories.  Some of the non-perturbative features of these new field theories
should be the ones we are describing in this paper.   

Aside from this and on a more geometrical note, we believe the backgrounds presented in
 this paper may serve as a prototype for  
new dynamical $SU(2)$ solutions of massive IIA supergravity that will be the 
corresponding string duals to the new field theories described above. 
This is, of course, in the same vein as the route from the conformal 
geometry of Klebanov-Witten to the non-conformal geometry of Klebanov-Strassler.       

In our view, these represent the most interesting avenues of further investigation.

\section*{Acknowledgments}
  
Discussions with various colleagues helped to improve the contents and  presentation of this paper.
We wish to thank: Ofer Aharony, Oren Bergman, Ben Craps, Tim Hollowood,  Carlos Hoyos, Georgios Itsios,  Zohar 
Komardgoski, Yolanda Lozano, Luca Martucci, Eoin O'Colgain,  Diego Rodriguez-Gomez, 
Daniel Schofield, Kostas Sfetsos, Jacob Sonnenschein, and Brian Wecht.
The work of J.G. was funded by the DOE Grant DE-FG02-95ER40896. N. Macpherson is supported by an STFC studentship.
Carlos Nunez is a Feinberg Foundation Visiting Faculty Program Fellow, he thanks the hospitality extended at Weizmann Institute and The Academic Study Group for the Isaiah Berlin Travel award.   D. Thompson is supported in part by the Belgian Federal Science Policy
Office through the Interuniversity Attraction Pole P7/37, and in part by the
``FWO-Vlaanderen" through the project G.0114.10N and through an ``FWO-Vlaanderen"
postdoctoral fellowship project number 1.2.D12.12N.

\appendix
\renewcommand{\thesection}{\Alph{section}}
\renewcommand{\theequation}{\Alph{section}.\arabic{equation}}

 \addtocontents{toc}{\protect\setcounter{tocdepth}{1}}  

\section{Conventions: Supergravity and G-structures}
\label{Appendix1conventions}
\setcounter{equation}{0}
\subsection{Supergravity}
We work in string and the 10-d hodge dual is defined such that
\be
F_n=(-1)^{int[n/2]}\star F_{10-n}.
\eeq
where $F_n$ are the RR fluxes of either type-IIA or type-IIB supergravity. The fluxes may be used to define a polyform F such 
that
\beq
F=\left\{\begin{array}{l l }
F_0+ F_2 +F_4 + F_6 + F_8 + F_{10} &~~~~\text{ for Type-IIA}\\
F_1+ F_3 +F_5 + F_7 + F_9 &~~~~\text{ for Type-IIB}\\
\end{array}\right. .
\eeq
In terms of the polyform the Bianchi identities may be expressed as
\beq
(d-H\wedge)F = 0,~~~ dH=0.
\eeq
It is easy to show this is satisfied with the definition
\beq
F=(d-H\wedge)C+F_0 e^{B_2}
\eeq
where $C$ is a polyform constructed from the RR potentials in the same fashion as above and $F_0$ should be taken to be 
zero in type-IIB. The flux equations of motion are expressed as
\beq
(d+H\wedge)\star F=0,~~~~ d(e^{-2\Phi} \star H)= \frac{1}{2} \sum_n F_{n}\wedge \star F_n.
\eeq
where the sum needs to me taken over the appropriate RR fluxes of type-IIA/IIB.

The dilaton must obey the equation of motion
\beq
d\star d\Phi+\star\frac{R}{4}-d\Phi\wedge\star d\Phi-\frac{1}{8}H\wedge \star H=0,
\eeq
while Einstein's equations are in type-IIA by
\beq
R_{\mu\nu}=-2D_{\mu}D_{\nu}\hat{\Phi}+\frac{1}{4}H^2_{\mu\nu}+e^{2\Phi}\bigg[\frac{1}{2} (F_2^2)_{\mu\nu}+\frac{1}{12}
(F_4^2)_{\mu\nu}-\frac{1}{4}g_{\mu\nu}(F_0^2+\frac{1}{2}F_2^2+\frac{1}{4!}F_4^2)\bigg],
\eeq
with an equivalent equation holding in type-IIB.

\subsection{Pure Spinors} \label{sec: spinorconvensions}
Here we follow the conventions of \cite{Andriot} except for a difference in the self duality condition of the RR section which leads to a few sign differences. We work in string frame and consider solution with metrics that can be expressed as
\begin{equation}
ds^2 = e^{2 A } dx^2_{3,1}+  ds^2_6 \ 
\end{equation}
and preserve $\mathcal{N}=1$ SUSY in 4-d with non trivial RR sector. This means that the internal space, with metric 
$ds^2_6$, must support an $SU(3)\times SU(3)$-structure \cite{Martucci:2005ht}. We decompose the 10-d MW spinors into a 
$4+6$ split as
\beq
\epsilon^1=\xi_+\otimes \eta^1_+ +\xi_-\otimes \eta^1_-,~~~ \epsilon^2=\xi_+\otimes \eta^2_{\mp} +\xi_-\otimes \eta^2_{\pm}.
\eeq
where in $\epsilon_2$ the upper/lower signs should be taken in type-IIA/B, the $\pm$ indicates chirality of both 4-d and 
internal 6-d spinors and we choose a basis for the internal spinors such that $(\eta_+)^*=\eta_-$. It is possible to define two 
$Cliff(6,6)$ pure spinors on the internal space as
\beq
\Psi_{\pm}=\eta^1_+\otimes (\eta^2_{\pm})^{\text{\textdagger}}
\eeq
which may be identified with polyforms under the Clifford map. The internal spinors are decomposed as
\beq
\eta^1_{+}= e^A e^{i\frac{\theta_++\theta_-}{2}}\eta_{+},~~~\eta^2_{+}= e^A e^{-i\frac{\theta_+-\theta_-}{2}}(k_{||}\eta_{+}
+k_{\perp}\chi_+)
\eeq
where $k_{||}^2+k_{\perp}^2=1$, $\eta_+^{\text{\textdagger}}\eta_+= \chi_+^{\text{\textdagger}}\chi_+=1$ and $\chi_
+^{\text{\textdagger}}\eta_+=0$.
The $\mathcal{N}=1$ SUSY conditions for such a $SU(3)\times SU(3)$-structure solution are given by the differential 
conditions
\beq
\begin{array}{ll}
\vspace{3 mm}
&(d-H\wedge)(e^{2A-\phi}\Psi_{\pm})=0\\
&(d-H\wedge)(e^{2A-\phi}\Psi_{\mp})=e^{2A-\phi}dA\wedge\bar{\Psi}_2\mp\frac{1}{8}e^{3A}\star_6 i\l(\tilde{F})
\end{array}
\eeq
where $\l(A_n)= (-1)^{\frac{n(n-1)}{2}}A_n$ and $\tilde{F}$ is the internal part of RR polyform in type IIA/B where the RR forms 
are each decomposed such that
\beq
F_n = \tilde{F}_n \mp e^{4A}vol_4\wedge \l(\star_6 \tilde{F}_{10-n}).
\eeq
As before upper/lower signs correspond to type IIA/B

Clearly in general $\eta^2_{+}$ is composed of a parts that is parallel and a part that is orthogonal to $\eta^1_{+}$. The 
$SU(3)\times SU(3)$-structure can categorised into 3 distinct cases depending on the values of the coefficients $k_{\perp}$ 
and $k_{||}$:

\subsection*{\textit{ $SU(3)$-structure}}
When $k_{\perp}=0$ the internal spinors are parallel and the pure spinors define an $SU(3)$-structure in 6-d such that
\beq
\begin{split}
\Psi_+ &=-e^{i\theta_+}\frac{e^A}{8}e^{-i J},\\
\Psi_- &=-ie^{i\theta_-}\frac{e^A}{8}\Omega_{hol}
\end{split}
\eeq 
where $J$ and $\Omega_{hol}$ are the two and holomorphic three forms associated with $SU(3)$, they are defined as in 
terms of the 6-d gamma matrices as
\beq\label{eq: JOmegadefs}
\Omega^{(hol)}_{abc}=-i \eta_-^{\text{\textdagger}} \gamma_{abc} \eta_+,~~~ J_{ab}=-i \eta_+^{\text{\textdagger}}
\gamma_{ab}\eta_+,
\eeq
and satisfy
\beq
J\wedge\Omega_{hol}=0,~~~J\wedge J\wedge J= \frac{3 i}{4} \Omega_{hol}\wedge \bar{\Omega}_{hol}.
\eeq
\subsection*{\textit{Orthogonal $SU(2)$-structure}}
When $k_{||}=0$ the internal spinors are orthogonal and the pure spinors define an orthogonal $SU(2)$-structure in 6-d such 
that
\beq
\begin{split}
\Psi_+ &=-i e^{i\theta_+}\frac{e^A}{8}e^{-v\wedge w}\wedge\omega,\\
\Psi_- &=ie^{i\theta_-}\frac{e^A}{8}(v+i w)\wedge e^{-i j}
\end{split}
\eeq
where the $SU(2)$-structure one forms $v$, $w$ and two forms $j$, $\omega$ are defined as
\beq\label{eq: su2forms}
w_a-i v_a =\eta_-^{\text{\textdagger}} \gamma_{a} \chi_+,~~~j_{ab}=-i\eta_+^{\text{\textdagger}} \gamma_{ab}\eta_+ +i\chi_
+^{\text{\textdagger}}\gamma_{ab}\chi_+,~~~ \omega_{ab}=\eta_-^{\text{\textdagger}}\gamma_{ab}\chi_-.
\eeq
and obey the relations
\beq\label{eq: su2str}
\begin{split}
&j\wedge \omega=\omega\wedge \omega =\iota_{(w-i v)}(\omega) = \iota_{(w-i v)} (j)=0\\
&j\wedge j= \frac{1}{2} \omega \wedge \bar{\omega}.
\end{split}
\eeq
\subsection*{\textit{Intermediate and Dynamical $SU(2)$-structure}}
For intermediate $SU(2)$-structure $k_{||}$ and $k_{\perp}$ are non zero constants, this and the previous example are also 
referred to as static $SU(2)$-structure. For dynamical $SU(2)$-structure  $k_{||}$ and $k_{\perp}$ are point dependent. For 
both these cases the pure spinors are given by
\beq
	\begin{aligned}
		\Phi_+ &= \frac{e^{A}}{8} e^{i \theta_+} e^{-i v\wedge w} \big( k_{\|} e^{-ij} - i k_{\perp} \omega \big) \\
		\Phi_- &= \frac{i e^{A}}{8} e^{i\theta_-} (v+i w)\wedge \big( k_{\perp} e^{-ij} + i k_{\|} \omega \big)
	\end{aligned},
\eeq
where eq \ref{eq: su2str} and eq \ref{eq: su2forms} still hold.

In these conventions the SUSY conditions (here we consider type IIA, details of type IIB are given in appendix E)  may be split up as follows:
\beq
\begin{array}{ll}
\vspace{3 mm}
&d\big[e^{3A-\hat{\Phi}} k_{\|}\big]=0\\
\vspace{3 mm}
&d\big[e^{3A-\hat{\Phi}}\big(k_{\|}(j+v\wedge w)+k_{\perp}\omega)\big)\big]-i e^{3A-\hat{\Phi}}k_{\|}H=0\\
\vspace{3 mm}
&d\big[e^{3A-\hat{\Phi}}\big(\frac{1}{2}k_{\|}(j+v\wedge w)^2+k_{\perp}v\wedge w \wedge\omega\big)\big]-i e^{3A-\hat{\Phi}}H
\wedge\big(k_{\|}(j+v\wedge w)+k_{\perp}\omega\big)=0\\
\vspace{3 mm}
\label{eq: oddforms}\end{array}
\eeq
where the second of these gives a definition for $H$ which can be combined with the first to give a definition of the NS 
potential, namely
\beq
B_2=-\frac{k_{\perp}}{k_{\|}}Im \omega
\eeq
this is not the same as the NS potential generated by non-abelian T-duality but must match it up to an exact.

The rest of the SUSY conditions are
\beq
\begin{array}{ll}
\vspace{3 mm}
&\star_6F_6=0\\
\vspace{4 mm}
&d\big[e^{4A-\hat{\Phi}}k_{\perp}\big(\sin\theta_-w-\cos\theta_-v\big)\big]=-e^{4A}\star_6F_4\\
\vspace{4 mm}
&d\big[e^{2A-\hat{\Phi}}k_{\perp}\big(\sin\theta_- v+\cos\theta_-w\big)\big]=0\\
\vspace{4 mm}
&d\big[e^{4A-\hat{\Phi}}\big(k_{\|}(\sin\theta_-\text{Im}\omega-
\cos\theta_-\text{Re}\omega)\wedge w-k_{\|}
(\sin\theta_-\text{Re}\omega+\cos\theta_-\text{Im}\omega)\wedge v+\\
\vspace{4 mm}
&~~~ k_{\perp}(\sin\theta_- v+\cos\theta_- w)\wedge j\big)\big]+ e^{4A-\hat{\Phi}}k_{\perp}H\wedge(\sin\theta_- w -\cos\theta_- 
v)=-e^{4A}\star_6F_2\\
\vspace{5 mm}
&d\big[e^{2A-\hat{\Phi}}\big(k_{\|}(\sin\theta_- \text{Re}\omega+\cos\theta_- \text{Im}\omega)\wedge w-k_{\|}(\cos\theta_- 
\text{Re}\omega
-\sin\theta_- \text{Im}\omega)\wedge v -\\
\vspace{3 mm}
&~~~ k_{\perp}(\sin\theta_- w -\cos\theta_-v)\wedge j\big)\big]+ k_{\perp} e^{2A-\hat{\Phi}}H\wedge(\cos\theta_- w+\sin\theta_- 
v)=0\\
\vspace{5 mm}
&d\big[\frac{1}{2}e^{4A-\hat{\Phi}}k_{\perp} j\wedge j\wedge(\cos\theta_- v-\sin\theta_- w)\big]+
e^{4A-\hat{\Phi}}H\wedge\big(k_{\|}(\sin\theta_-\text{Im}\omega-\cos\theta_-\text{Re}\omega)\wedge w-\\
\vspace{3 mm}
&~~~k_{\|}(\sin\theta_-\text{Re}\omega+\cos\theta_-\text{Im}\omega)\wedge v+k_{\perp}(\sin\theta_- v+\cos\theta_- w)\wedge j
\big)=-e^{4A}\star_6F_0\\
\vspace{5 mm}
&d\big[\frac{1}{2}e^{2A-\hat{\Phi}}k_{\perp} j\wedge j\wedge(\cos\theta_- w+\sin\theta_- v)\big]+ e^{2A-\hat{\Phi}}H\wedge
\big(-k_{\|}
(\sin\theta_- \text{Re}\omega+\cos\theta_- \text{Im}\omega)\wedge w+\\
\vspace{3 mm}
&~~~k_{\|}(\cos\theta_- \text{Re}\omega-\sin\theta_- \text{Im}\omega)\wedge v + k_{\perp}(\sin\theta_- w -\cos\theta_-v)\wedge 
j\big)=0\\
\end{array}
\label{eq: evenforms}\eeq
from which it is possible to define the higher forms of the RR sector as:
\beq
\begin{array}{ll}
\vspace{3 mm}
F_6&=dC_5\\
\vspace{3 mm}
F_8&= dC_7-H\wedge C_5\\
\vspace{3 mm}
F_{10}&= dC_9-H\wedge C_7
\end{array}
\eeq
where the RR potentials are given by:
\beq
\begin{array}{ll}
\vspace{3 mm}
C_5&=e^{4A-\hat{\Phi}}vol_4\wedge k_{\perp}\big(\sin\theta_-w-\cos\theta_-v\big)\\
\vspace{3 mm}
C_7&=-e^{4A-\hat{\Phi}}vol_4\wedge\bigg[k_{\|}(\sin\theta_-\text{Im}\omega-
\cos\theta_-\text{Re}\omega)\wedge w-\\
\vspace{4 mm}
&~~~~k_{\|}(\sin\theta_-\text{Re}\omega+\cos\theta_-\text{Im}\omega)\wedge v+k_{\perp}(\sin\theta_- v+\cos\theta_- w)\wedge 
j\bigg]\\
\vspace{3 mm}
C_9&=\frac{1}{2}e^{4A-\hat{\Phi}}vol_4\wedge k_{\perp} j\wedge j\wedge(\cos\theta_- v-\sin\theta_- w)\\
\end{array}
\eeq
The calibration is given by
\beq
\Psi_{cal}=-8 e^{3A-\hat{\Phi}}Im \Phi_- e^{\pm B_2}
\eeq
where $\pm$ depends on our conventions in the WZ action. That $S_{DBI}+S_{WZ}=0$ is trivial because in these 
convensions we have:
\beq
C_5+C_7+C_9=-8vol_4\wedge e^{3A-\hat{\Phi}}Im \Phi_-
\eeq
This all works perfectly for the case $\theta_-=0$ which is the dual of the wrapped D5 solution.

\section{Details of the non-Abelian T-duality on the 
D5 branes solution.}\label{detailsD5}
The purpose of this section is to give some details of the $SU(2)$ isometry T-dual of Wrapped D5 branes on $S^2$. This was 
first derived in \cite{Itsios:2013wd}, but in slightly different conventions and the G-structure was not found. This is the $
\mathcal{C}=1, ~\mathcal{S}=0$ limit of the full baryonic branch dual solution, and as the procedure for find the the G-structure 
is the same in both case we hope that this more simple example will be instructive.

Solution of wrapped D5 branes on $S^2$ \cite{Casero:2006pt} has string frame metric given by
\beq
\begin{split}
ds^2=& e^{\Phi}\bigg( dx_{1,3}^2+ e^{2k}d\r + e^{2h}\big(d\theta^2+\sin^2\theta d\varphi^2\big)+\\
&~~~~~~ \frac{e^{2g}}{4}\big((\tilde{\o}_1+a d\theta)^2+(\tilde{\o}_2-a \sin\theta d\varphi)^2\big)+ \frac{e^{2k}}{4} (\tilde{\o}_3+
\cos d\varphi)^2\bigg)
\end{split}
\eeq
where the functions $a,b,g,h,k$ and the dilaton $\Phi$ only depend on the holographic coordinate $r$. The $\tilde{\o}_i$ are 
$SU(2)$ left invariant 1-forms which can be parametrised as
\beq
	\begin{aligned}\label{eq: leftInv}
		\tilde{\o}_1 & = \cos\psi d\tilde\theta + \sin\psi\sin\tilde\theta d\tilde\varphi \, , \\
		\tilde{\o}_2 & = -\sin\psi d\tilde\theta +\cos\psi\sin\tilde\theta d\tilde\varphi \, , \\
		\tilde{\o}_3 & = d\psi + \cos\tilde\theta d\tilde\varphi \, .
	\end{aligned}.
\eeq
A convenient set of vielbeins is given by
\begin{align}
	e^{x^i}			&= e^{\frac{\Phi}{2}}
dx^i    
\,,\;\;\;
	e^{\r}	 		=  e^{\frac{\Phi}{2}+k} 
d\r  
\,,\;\;\;
	e^{\theta}	=  e^{\frac{\Phi}{2}+h} 
    d\theta  
\,,\;\;\;
	 e^{\varphi}= e^{\frac{\Phi}{2}+h} 
 	\sin\theta \,
d\varphi\,    ,\nonumber\\
	e^{1}				&=  \frac{1}{2}e^{\frac{\Phi}{2}+g} 
(\tilde{\omega}_1 +a\, d\theta)\,  ,\qquad\qquad
	e^{2}				=		\frac{1}{2}
e^{\frac{\Phi}{2}+g} 
	(\tilde{\omega}_2 	
-a\,\sin\theta\, d\varphi)   
\,,\nonumber\\ 
	e^{3}				&= \frac{1}{2}e^{\frac{\Phi}{2}+k} 
(\tilde{\omega}_3 +\cos\theta\, d\varphi)\,.
	\label{vielbeinD5}
\end{align}
with respect to which the non trivial RR flux $F_3$ may be expressed as
\beq
F_3= e^{-\frac{3}{2}\Phi}
\Big[f_1 e^{123}+ f_2 e^{\theta\varphi 3}
+ f_3(e^{\theta23}+ e^{\varphi 13})+ 
f_4(e^{\r 1\theta}+ e^{\r\varphi 2})   \Big]
\eeq
where the $f_i$ are given by eq \ref{eq: fidef}. In these conventions the projections the 10-d Killing spinor $\epsilon$ obeys 
are
\beq\label{eq: projectors}
\Gamma_{12}\epsilon=\Gamma_{\theta\varphi}\epsilon,~~~\Gamma_{r123}\epsilon= (\cos\alpha+\sin\alpha \Gamma_{\varphi 
2})\epsilon, ~~~ i\epsilon^*=\epsilon \ ,
\eeq
with respect to the $4+6$ split we can define components of $\epsilon$ to be equal with positive chirality as
\beq
\epsilon_{1}=\epsilon_{2}= e^{A}(\xi_{+}\otimes \eta_++\xi_{-}\otimes \eta_-)
\eeq
where $2A=\Phi$. Once the usual decomposition of gamma matrices,
\beq \label{eq: gammade}
\Gamma_{\mu}= \hat{\gamma}_\mu\otimes \mathbb{I},~~~~ \Gamma_{a}= \mathbb{I}\otimes \gamma_{a}
\eeq
is performed it is a simple matter to derive the $SU(3)$-structure forms of eq \ref{eq: JJSU3} using eq \ref{eq: JOmegadefs}, 
where we have chosen $i\gamma_{r\theta\varphi123}\eta_+=\eta_+$. To do this it is helpful to perform a rotation in 
$e^{\varphi}$, $e^{2}$ which will also be useful later
\beq\label{eq: CanonicalRot}
	\begin{aligned}
		\hat e^{\varphi} &= \cos \a e^\varphi + \sin \a e^2 \\
		\hat e^2 &= -\sin \a e^\varphi + \cos \a e^2 \\
		\hat e^a &= e^a \text{ for } a \neq \varphi, 2.
	\end{aligned}
\eeq

The rotated 6-d projections are then simply
\beq \label{eq:projj}
	\hat \gamma_{\varphi \theta} \eta_+ = \hat \gamma_{r3} \eta_+ = \hat \gamma_{2 1} \eta_+ = i \, \eta_+
\eeq
and the $SU(3)$-structure becomes canonical.

We want to T-dualise this wrapped D5-brane solution along the $SU(2)$ isometry parametrised by $\tilde{\omega_i}$. 
Section 2 and Appendix B of \cite{Itsios:2013wd} give all the details of the algorithm one must follow to do this and so we 
direct the interested reader there for details of the NS sector. For the RR sector we only give details that will be relevant for 
later calculations.

The duality will drastically change the vielbeins that contain the $SU(2)$ left invariant 1-forms $e^1$, $e^{2}$, $e^3$ and 
leave the others untouched. For the dual of the wrapped D5 brane solution gauge fixed such that the remaining dual 
coordinates are $v_2$, $v_3$ and $\psi$, the cannonical vielbeins given by the procedure of \cite{Itsios:2013wd} are
\beq
\begin{array}{ll}\label{eq: dualveildash}
e^{\hat{1}'}&=\frac{e^{g+\frac{\Phi }{2}}}{8 {\cal W}} \bigg[e^{2 k+\Phi } \bigg(-\sqrt{2} e^{2 g+\Phi }
   (\cos \psi (a \o_2 v_3+dv_2)+\sin \psi (a\o_1 v_3-\o_3 v_2))-\\
   &~~~~~~~~~~~~4 v_3 \sin\psi (a \o_2 v_3+dv_2)+4 v_3 \cos \psi(a \o_1 v_3-\o_3 v_2)\bigg)-\\
   &~~~~~~~~~~~~4 v_2e^{2 g+\Phi } \sin \psi (a \o_2 v_2-dv_3)-8 \sqrt{2} v_2 \cos \psi (v_2 dv_2+v_3dv_3)\bigg]\\
e^{\hat{2}'}&=\frac{e^{g+\frac{\Phi }{2}}}{8 {\cal W}} \bigg[e^{2 k+\Phi } \bigg(\sqrt{2} e^{2 g+\Phi } (\cos
   \psi (\o_3 v_2-a \o_1 v_3)+a\o_2 v_3 \sin \psi+dv_2 \sin \psi)-\\
   &~~~~~~~~~~~~4 v_3(\cos \psi (a \o_2 v_3+dv_2)+\sin \psi (a\o_1 v_3-\o_3 v_2))\bigg)-\\
   &~~~~~~~~~~~~4 v_2e^{2 g+\Phi } \cos \psi (a \o_2 v_2-dv_3)+8\sqrt{2} v_2 \sin \psi (v_2 dv_2+v_3dv_3)\bigg]\\
e^{\hat{3}'}&=\frac{e^{k+\frac{\Phi }{2}} }{8 {\cal W}}\bigg[\sqrt{2} e^{4 g+2 \Phi } (a \o_2v_2-dv_3)+4 v_2 e^{2 g+\Phi } 
(\o_3v_2-a \o_1 v_3)-\\
	 &~~~~~~~~~~~~8 \sqrt{2} v_3 (v_2 dv_2+v_3 dv_3)\bigg]\\
\end{array}
\eeq
with the remaining veilbeins still given by eq \ref{vielbeinD5}, that is $e^{a'}=e^{a}$ for $a\neq 1,2,3$. The $\o_i$ are defined 
as in eq \ref{eq: leftInv} but with $\tilde{\theta}\to\theta,~\tilde{\varphi}\to\varphi$. It is possible to remove all the explicit angular 
dependence from the dual solution by performing a rotation in the $\theta$, $\varphi$ directions such that
\beq
\begin{split}
e^{\hat{\theta}}&= e^{h+\Phi/2} \o_1 = \cos\psi e^{\theta} + \sin\psi e^{\varphi}\\
e^{\hat{\varphi}}&= e^{h+\Phi/2} \o_2 = -\sin\psi e^{\theta} + \cos\psi e^{\varphi} \ , 
\end{split}
\eeq
and an additional rotation in 1',2',3' directions such that
\beq\label{eq: psirot} 
\begin{split}
e^{\hat{1}}&= \cos\psi e^{1'}-\sin \psi e^{2'}\\
e^{\hat{2}}&=\sin\psi e^{1'}+\cos \psi e^{2'}\\
e^{\hat{3}}&= e^{3'}.
\end{split}
\eeq
Theses rotation make the expressions for the vielbeins and fluxes a lot more simple than they otherwise would be, they are 
given for the dual of the wrapped D5 solution as in section \ref{NATDBB} but with $\mathcal{S}=0,~\mathcal{C}=1$. However, 
it is the $e^{a'}$ vielbeins rather than the $e^{\hat{a}}$ ones that are more suited to calculating the G-structure of the dual 
solution. 

It was shown explicitly in \cite{Barranco:2013fza} that the 10-d MW Killing spinors transform under an $SU(2)$ isometry T-duality as 
\beq\label{eq: D5dualKillingspinor}
\hat{\epsilon}_1=\epsilon_1,~~~\hat{\epsilon}_2=\Omega\epsilon_2.
\eeq
where $\Omega$ is given by
\beq\label{eq: D5Omega}
\Omega= \Gamma^{(10)}\frac{-\Gamma_{123} + \sum_{a=1}^3 \zeta_a\Gamma^a}{\sqrt{1+\zeta^2}}
\eeq
and for the wrapped D5 background we have
\beq
\zeta^1=2\sqrt{2}e^{-g-k-\phi}v_2\cos\psi,~~~\zeta^2=-2\sqrt{2}e^{-g-k-\phi}v_2\sin\psi,~~~\zeta^3=2\sqrt{2} e^{-2g-\phi}v_3.
\eeq
Starting from eq \ref{eq: dualveildash} we first rotate the veilbeins as in eq \ref{eq: CanonicalRot} so that the projections are 
canonical. The $\Omega$ matrix then becomes 
\beq
	\Omega = \frac{1}{\sqrt{1 + \zeta.\zeta}} \big( \cos \alpha \hat\Gamma^{123} + \sin \alpha \hat\Gamma^{1\varphi3} + 
\zeta_1 \hat\Gamma^1 + \zeta_2 \cos \alpha \hat\Gamma^2 + \zeta_2 \sin\alpha \hat\Gamma^{\varphi} + \zeta_3 \hat
\Gamma^3 \big)
\eeq
where we have used $\gamma^{1\varphi3}\eta_+=i\eta_-$. The new spinor $\hat \epsilon_2$ is:
\beq
	\hat \epsilon_2 = e^{\Phi/4} \big( \z_+ \otimes \hat\eta^2_- + \z_- \otimes \hat\eta^2_+ \big)
\eeq
where
\beq\label{eq: etadualD5}
	\hat\eta^2_- = \frac{\cos \alpha \hat\gamma^r + \zeta_1 \hat \gamma^1 + \zeta_2 \cos\alpha \hat \gamma^2 + \zeta_3 \hat
\gamma^3 + \zeta_2 \sin\alpha \hat \gamma^\varphi}{\sqrt{1 + \zeta.\zeta}} \eta_+ + i \frac{\sin \alpha}{\sqrt{1 + \zeta.\zeta}} 
\eta_-.
\eeq
It is clear here that, as long as $\sin \alpha \neq 0$, we are in the dynamical $SU(2)$-structure case, because $\alpha=
\alpha(r)$.
In order to simplify the expressions we perform another transformation of the vielbein basis:
\beq\label{eq:RmatixD5}
	R = \frac{1}{\sqrt{\Delta}}\left(
	\begin{array}{cccccc}
		\cos\alpha & 0 & 0 & \zeta_1 & \zeta_2\cos\alpha & \zeta_3 \\
		0 & \sqrt{\Delta} & 0 & 0 & 0 & 0 \\
		0 & 0 & \sqrt{\Delta} & 0 & 0 & 0 \\
		-\zeta_1 & 0 & 0 & \cos\alpha & \zeta_3 & -\zeta_2\cos\alpha \\
		-\zeta_2\cos\alpha & 0 & 0 & -\zeta_3 & \cos\alpha & \zeta_1 \\
		-\zeta_3 & 0 & 0 & \zeta^2\cos\alpha & -\zeta^1 & \cos\alpha \\
	\end{array}
	\right)
\eeq
where
\beq
	\Delta = \cos^2 \alpha+\zeta_1^2+\zeta_2^2 \cos^2\alpha+\zeta^2_3
\eeq
We define a new basis:
\beq
	\tilde e = R.\hat e
\eeq
where the order is $r\theta\varphi123$. In terms of this new basis, the spinor is:
\beq
	\tilde \eta^2_- = \left(\frac{\sqrt{\Delta} \tilde \gamma^r + \zeta_2 \sin\alpha \tilde \gamma^\varphi}{\sqrt{1+\zeta.\zeta}} 
\right) \eta_+ + i \frac{\sin \alpha}{\sqrt{1 + \zeta.\zeta}} \eta_-
\eeq
And the projections in this basis are still:
\beq \label{eq:proj2}
	\tilde \gamma_{\varphi \theta} \eta_+ = \tilde \gamma_{r3} \eta_+ = \tilde \gamma_{2 1} \eta_+ = i \, \eta_+
\eeq
Let us now express the forms of the geometric structure, following the conventions of appendix \ref{Appendix1conventions}.
\beq
	\begin{aligned}
		&e^{2A} = e^{\Phi} \\
		&\theta_+ = 0 \qquad \theta_- = 0 \\
		&k_{\|} = \frac{\sin \alpha}{\sqrt{1+\zeta.\zeta}} \qquad k_{\perp} = \sqrt{\frac{\cos^2 \alpha + \zeta.\zeta}{1+\zeta.
\zeta}} \\
		&z = w - i \, v = \frac{1}{\sqrt{\cos^2 \alpha + \zeta.\zeta}} \big( \sqrt{\Delta} \tilde e^3 + \zeta_2 \sin\alpha \tilde e^
\theta + i (\sqrt{\Delta} \tilde e^r + \zeta_2 \sin \alpha \tilde e^{\varphi}) \big) \\
		&j = \tilde e^{r3} + \tilde e^{\varphi \theta} + \tilde e^{21} - v \wedge w \\
		&\omega =  \frac{i}{\sqrt{\cos^2 \alpha + \zeta.\zeta}} \big( \sqrt{\Delta} (\tilde e^\varphi+ i \tilde e^\theta) - \zeta_2 \sin
\alpha (\tilde e^r + i \tilde e^3) \big) \wedge (\tilde e^2 + i \tilde e^1)
	\end{aligned}
\eeq
which is a dynamical $SU(2)$-structure.
\section{ Details of the non-Abelian 
T-duality on the Baryonic Branch solution}\label{detailsBB}
In this section we give some details of the $SU(2)$ isometry T-dual of the Baryonic branch of Klebanov-Strassler. This was originally derived in \cite{Itsios:2013wd} with gauge fixing such that $v_1=\varphi=\theta=0$. The previous derivation indicated a departure in the T-dual from the log corrected $AdS_5$ asymptotics of the baryonic branch. Let as begin by giving some details of original calculation in our current convensions

\subsection{Dual of the baryonic branch \textbf{without} the shift in $B_2$}\label{sec: BB no B2}
Once more we will start by specifying the dual vielbeins.
The components 
\begin{align}
	e^{x^i}			&= e^{\frac{\Phi}{2}}
\hat{h}^{-\frac{1}{4}} dx^i    
\,,\;\;\;
	e^{\r}	 		=  e^{\frac{\Phi}{2}+k} 
\hat{h}^{\frac{1}{4}}d\r  
\end{align}
do not change. The vielbeins in the $\theta,\varphi$ are also unchanged by the duality however we find it useful to introduce a 
rotation in $e^{\theta},e^{\varphi}$ such that the dual solution has no explicit $\psi$ dependence.
\beq
e^{\hat{\theta}}= \sqrt{\mathcal{C}}e^{h+\Phi/2} 
\o_1,~~~~ e^{\hat{\varphi}}=\sqrt{\mathcal{C}}e^{h+\Phi/2}\o_2,
\eeq
The vielbeins in the directions $\hat{1},\hat{2}, \hat{3}$ 
can be compactly written in terms of the quantities defined as,
\begin{equation}
\begin{split}
\mathcal{V}_3&= v_3+ \frac{e^{2g+\Phi}}{2\sqrt{2}}\mathcal{S} 
\cos\alpha  \ ,  \\[3 mm]
\Lambda&= d \mathcal{V}_3 +\frac{e^{\Phi-2h}}{2 \sqrt{2}}
\mathcal{S}N_c\bigg(e^{2g}+2e^{2h}-ae^{g}(be^{g}-2 e^{h}\cot \alpha)\bigg)
d\r  \ ,  \\[3 mm]
\mu_1&= a e^g \cos\alpha +2e^h \sin\alpha \ , \\[3 mm]
\label{eq: simpnoB2}
\end{split}
\eeq
With these, we have
\beq
\begin{split}\label{eq :BBDualVeil2}
e^{\hat{1}}&=\frac{e^{g+\Phi/2}}{16 \mathcal{W}}\sqrt{\mathcal{C}}
\bigg[e^{2k+\Phi}\bigg(8 \mathcal{V}_3(a\mathcal{V}_3\o_1-v_2\o_3)-2
\sqrt{2}e^{2g+\Phi} 
\mathcal{C}(dv_2+a\mathcal{V}_3 \o_2)-2\sqrt{2}e^{g+\Phi}\mathcal{S} 
\mathcal{V}_3 \mu_1 \o_1 \\
&~~~~~~~~~~~~~~~~~~~~~~~+e^{3g+2\Phi} \mathcal{C}\mathcal{S} 
\mu_1 \o_2\bigg)+
8v_2 \big(e^{g+\Phi}v_2 \mathcal{S}\mu_1\o_2-2\sqrt{2}
(\mathcal{V}_3 \Lambda+v_2 dv_2)\big)\bigg]\ , \\[3 mm]
e^{\hat{2}}&=\frac{e^{g+\Phi/2}}{16 \mathcal{W}}\mathcal{C}^{3/2}
\bigg[e^{2k}\bigg(-2\sqrt{2}e^{2g+\Phi}\mathcal{C}
(a \mathcal{V}_3\o_1-v_2 \o_3)-8 
\mathcal{V}_3(dv_2+a\mathcal{V}_3\o_2)+e^{3g+2\Phi}\mathcal{C}
\mathcal{S}\mu_1 \o_1\\
&~~~~~~~~~~~~~~~~~~~~~~~+2\sqrt{2}e^{g+\Phi}\mathcal{S}
\mathcal{V}_3\mu_1 \o_2\bigg)-
8 e^{2g}v_2(-\Lambda+av_2\o_2)\bigg] \ , \\[3 mm]
e^{\hat{3}}&=\frac{e^{k+\Phi/2}}{16 \mathcal{W}}\sqrt{\mathcal{C}}
\bigg[e^{g+\Phi}v_2\big(\sqrt{2}e^{2g+\Phi}
\mathcal{C}(ae^{g} \mathcal{C}\o_2+\mathcal{S}\mu_1 \o_1)-4e^{g} 
\mathcal{C}(a \mathcal{V}_3 \o_1-v_2 \o_3)+4 
\mathcal{S}\mathcal{V}_3\mu_1 \o_2)\\
&~~~~~~~~~~~~~~~~~~~~~~~-\sqrt{2}\Lambda(e^{4g+2\Phi} 
\mathcal{C}^2+8 \mathcal{V}_3^2)-8 \sqrt{2}v_2 \mathcal{V}_3dv_2\bigg] \ .
\end{split}
\eeq
where the rotation of eq \ref{eq: psirot} has been performed \footnote{ Actually this differs from \cite{Itsios:2013wd} in orientation which can be compensated for via $\hat{1}\leftrightarrow \hat{2}$.}.
We will then have a metric that in terms of these vielbeins reads,
$ds_{st}^2=\sum_{i=1}^{10} (e^{i})^2$.
Notice that the quantity $\Lambda$ in eq.(\ref{definitionsz}) will, when 
squared to construct the metric
with the vielbeins above, imply the existence of crossed terms
$g_{\r v_3}$ and also the change of the asymptotic behaviour of $g_{\r\r}$ away from $\log$ corrected $AdS_5$.

In terms of these vielbeins, the NS two-form $B_2$ reads,
\beq
\begin{split}\label{eq: BBB22}
\widehat{B}_2  = -& \frac{1}{4v_2}
\bigg(2e^{-h}a(e^gv_2 e^{\hat{\theta}\hat{1}}+e^{k}
\mathcal{V}_3 e^{\hat{\theta}\hat{3}})-4e^{k-g}
\mathcal{V}_3 e^{\hat{1}\hat{3}}+\sqrt{2}
\mathcal{C}e^{g+k+\Phi}e^{\hat{2}\hat{3}}\bigg)+\\
&~~~ \frac{\mathcal{S}}{\mathcal{C}}\bigg[\frac{\mathcal{V}_3e^k}{2v_2}
\big(ae^{-h} e^{r\hat{\theta}}-2 e^{-g} e^{r \hat{1}}\big)+
\frac{e^{g+k+\Phi-h}}{4 \sqrt{2} V_2}\mathcal{C} 
\big(2 e^{2h}e^{r \hat{2}} +\mu_1e^{\hat{\theta}\hat{1}}\big)-\\
& ~~~~~~~~~~~~~~~~~~~~~~~~\frac{e^{-h}}{2} \big(2 e^{h}\cos\alpha -
ae^{g}\sin\alpha \big)e^{\hat{\theta}\hat{\varphi}}+
e^{\r \hat{3}}-\frac{e^{-h}}{2}\mu_1 e^{\hat{\theta}\hat{2}}\bigg].
\end{split}
\eeq
The dual dilaton is given by
\be\label{eq: BBdil2}
\widehat{\Phi} =  \Phi - \frac{1}{2} \ln {\cal W}\ ,\;\;\;
\mathcal{ W} = \mathcal{C}\bigg(\frac{1}{8} e^{4g+2k+3\Phi}
\mathcal{C}^2 + e^{2g+\Phi}v_2^2 + e^{2k+\Phi}\mathcal{V}_3^2\bigg) \ .
\ee
And the RR sector is given by,
\beq
\begin{split}
\vspace{3 mm}
F_0 &= \frac{N_c}{\sqrt{2}}\ , \\[3 mm]
\vspace{3 mm}
F_2 &= -\frac{e^{-\Phi}}{4}N_c\mathcal{C}\bigg[2e^{-2h}\big(1+a^2-2ab\big)
\mathcal{V}_3 e^{\hat{\theta}\hat{\varphi}} + 
e^{-g-h-k}\mathcal{C}\big(a-b\big)
\bigg( \sqrt{2}e^{2g+k+\Phi}\big(e^{\hat{\theta}\hat{1}}-
e^{\hat{\varphi}\hat{2}}\big)+\\
&~~~~~~~~~~~~4e^{k}\mathcal{V}_3
\big(e^{\hat{\theta}\hat{2}}-e^{\hat{\varphi}\hat{1}}\big)-
4v_2e^{g}e^{\hat{\varphi}\hat{3}}\bigg)-
8e^{-2g}\mathcal{V}_3 e^{\hat{1}\hat{2}}-8e^{-g-k}v_2e^{\hat{2}\hat{3}}
-2e^{-h-k}v_2e^{r\hat{\theta}}\bigg]-\\
&~~~~~~~\frac{\mathcal{S}e^{g-h}}{\sqrt{2}
\mathcal{C}\sin\alpha}\bigg(N_c b+ a(e^{2g}\cos^2\alpha-N_c)
+e^{g+h}\sin2\alpha\bigg)e^{\hat{\theta}\hat{\varphi}} \ ,  \\[3 mm]
F_4&=\frac{e^{-g-h-k-\Phi}}{8\mathcal{C}}N_c
\bigg[\mathcal{C}\big(1+a^2-2ab\big)e^{\hat{\theta}\hat{\varphi}}
\wedge\big(\sqrt{w}e^{2g+k+\Phi-h}
e^{\hat{1}\hat{2}}+4e^{2g-h}e^{\hat{1}\hat{3}}\big)\\
&~~~~~~~~~~~~~~~~~~~~~~~~\mathcal{C}b'e^{r\hat{\theta}}
\wedge\big(4e^{k}\mathcal{V}_3e^{\hat{1}\hat{3}}-
\sqrt{2}e^{2g+k+\Phi}e^{\hat{2}\hat{3}}\big)-8e^{g}v_2
\big(a-b\big)e^{\hat{\theta}\hat{1}\hat{2}\hat{3}}\\
&~~~~~~~~~~~~~~~~~~~~~~~~e^{r\hat{\varphi}}\wedge 
\big(4e^{g}v_2e^{\hat{1}\hat{2}}-b'e^k(\sqrt{2}e^{2g+\Phi}
e^{\hat{1}\hat{3}}+4\mathcal{V}_3 e^{\hat{2}\hat{3}})\big)\bigg]-\\
&~~~~~~~\frac{2\mathcal{S}e^{-g-h-k-\Phi}}{\mathcal{C}^2 \sin\alpha}
\bigg(a\big(e^{2g}\cos^2\alpha-N_c\big)+\big(N_c b+e^{g+h}
\sin2\alpha\big)\bigg)
\bigg(\mathcal{V}_3 e^{k}e^{\hat{\theta}\hat{\varphi}\hat{1}\hat{2}}
+v_2e^{g} e^{\hat{\theta}\hat{\varphi}\hat{2}\hat{3}}\bigg) \ . 
\end{split}
\eeq
We will now proceed to show that the bad asymptotic behaviour and off diagonal $\rho$ terms of the metric are actually a gauge artefact. 
\subsection{The dual of the baryonic branch \textbf{with} the shift in $B_2$}
The NS 2-from of the original solution contains the term
\beq
\tilde{B}_2=-\frac{1}{2} e^{2k+\Phi} \mathcal{S} (\tilde{\omega}_3+\cos\theta d\varphi)\wedge d\rho.
\eeq
It is this term, when dualised, that gives rise to the undesirable behaviour as this will contribute to the dual metric in both $g_{\rho\rho}$ and $g_{\rho v3}$ via the dual vielbeins $e^{\hat{i}}$ which will have legs in $\rho$. This happens because of the $d\rho\wedge\tilde{\o}_i$ term in $\tilde{B}_2$ which is not a spectator under the duality transformation\footnote{See section 2 of \cite{Itsios:2013wd} for details of how the initial $B_2$ enters into the definition of the dual vielbeins.}. However, one is always free to add an exact to the NS potential as this will not change the fluxes or metric of the original solution. Consider adding a closed form to the initial $B_2$
\beq
B_2\to B_2 + d\big(\mathcal{Z}(r)(\tilde{\omega}_3+\cos\theta d\varphi)\big)
\eeq 
This precisely cancels the effect of $\tilde{B}_2$ in the dual solution when  $\mathcal{Z}'=-\frac{1}{2}\mathcal{S}e^{2k+\Phi}$ because
\beq
\tilde{B}_2+d(\mathcal{Z}(r) \tilde{\omega}_3) =-\mathcal{Z}(\tilde\o_1\wedge\tilde\o_2+\sin\theta d\theta\wedge d\varphi)+\frac{1}{2}(\mathcal{S}e^{2k+\Phi}+2\mathcal{Z}')d\rho\wedge( \tilde\o_3+ \cos\theta d\varphi).
\eeq
As there is no longer a $d\rho\wedge\tilde{\o}_i$ term in the NS 2 form before dualisation, the dual vielbeins will have no legs in $\rho$ and so there will no longer be a modification to $g_{\rho\rho}$ and $g_{\rho v_3}$. The trade off is that the function $\mathcal{Z}$ will now enter into the dual solution.

We now once more follow the procedure of \cite{Itsios:2013wd} with gauge fixing, as before, such that $v_1=\varphi=\theta=0$. We are lead to the dual vielbeins
\beq
\begin{array}{ll}\label{eq: BBdualveildash}
e^{\hat{1}'}&=\frac{e^{g+\frac{\Phi }{2}}}{8 {\cal W}}\sqrt{\mathcal{C}} \bigg[e^{2 k+\Phi } \bigg(-\sqrt{2}\mathcal{C} e^{2 g+\Phi }
   (\cos \psi (a \o_2 \mathcal{H}+dv_2)+\sin \psi (a\o_1 \mathcal{H}-\o_3 v_2))-\\
   &~~~~~~~~~~~~~~~~~4 \mathcal{H} \sin\psi (a \o_2 \mathcal{H}+dv_2)+4 \mathcal{H} \cos \psi(a \o_1 \mathcal{H}-\o_3 v_2)\bigg)-\\
   &~~~~~~~~~~~~~~~~~4 v_2\mathcal{C}e^{2 g+\Phi } \sin \psi (a \o_2 v_2-dv_3)-8 \sqrt{2} v_2 \cos \psi (v_2 dv_2+ \mathcal{H}dv_3)+\\
   &~~~~~~~~~~~~~~~~~ \frac{1}{2}\mu_1\mathcal{S}e^{g+\Phi}\bigg(8v_2^2\cos\psi\o_2+\mathcal{C}e^{2k+\Phi}\big(\cos\psi(\mathcal{C}e^{2g+\Phi}\o_2-2\sqrt{2} \mathcal{H} \o_1)+\\ &~~~~~~~~~~~~~~~~~\sin\psi(\mathcal{C}e^{2g+\Phi}\o_1+2\sqrt{2}\mathcal{H}\o_2)\big)\bigg)\bigg]\\
e^{\hat{2}'}&=\frac{e^{g+\frac{\Phi }{2}}}{8 {\cal W}}\sqrt{\mathcal{C}} \bigg[e^{2 k+\Phi }\mathcal{C} \bigg(\sqrt{2}\mathcal{C} e^{2 g+\Phi } (\cos
   \psi (\o_3 v_2-a \o_1 \mathcal{H})+a\o_2 \mathcal{H} \sin \psi+dv_2 \sin \psi)-\\
   &~~~~~~~~~~~~~~~~~4 \mathcal{H}(\cos \psi (a \o_2 \mathcal{H}+dv_2)+\sin \psi (a\o_1 \mathcal{H}-\o_3 v_2))\bigg)-\\
   &~~~~~~~~~~~~~~~~~4 v_2\mathcal{C}e^{2 g+\Phi } \cos \psi (a \o_2 v_2-dv_3)+8\sqrt{2} v_2 \sin \psi (v_2 dv_2+ \mathcal{H}dv_3)+\\
   &~~~~~~~~~~~~~~~~~\frac{1}{2}\mu_1\mathcal{S}e^{g+\Phi}\bigg(-8v_2\sin\psi \o_2+ \mathcal{C}e^{2k+\Phi}\big( (\mathcal{C}e^{2g +\Phi}\o_1+2\sqrt{2}\mathcal{H}\o_2)\cos\psi-\\
&~~~~~~~~~~~~~~~~~(\mathcal{C}e^{2g+\Phi}\o_2-2\sqrt{2}\mathcal{H}\o_1)\sin\psi  \big)\bigg)\bigg]\\
e^{\hat{3}'}&=\frac{e^{k+\frac{\Phi }{2}} }{8 {\cal W}}\sqrt{\mathcal{C}}\bigg[\sqrt{2}\mathcal{C}^2 e^{4 g+2 \Phi } (a \o_2v_2-dv_3)+4 v_2\mathcal{C} e^{2 g+\Phi } 
(\o_3v_2-a \o_1 \mathcal{H})-\\
	 &~~~~~~~~~~~~~~~~~8 \sqrt{2} \mathcal{H} (v_2 dv_2+\mathcal{H} dv_3)+ \mu_1v_2\mathcal{S}e^{g+\Phi}(4\mathcal{H}\o_2+\sqrt{2} \mathcal{C}e^{2g+\Phi}\o_2)\bigg]\\
\end{array}
\eeq
which upon rotating according to eq \ref{eq: psirot} give the vielbeins of  eq \ref{eq :BBDualVeil}.

A valid question at this point is whether there is a local diffeomorphism which maps us from the baryonic branch dual solution as defined in section \ref{sec: BB no B2} to the solution defined as in section \ref{NATDBB}. The answer is yes, and it may be most easily found by comparing the dilaton as defined in eq \ref{eq: BBdil} and eq \ref{eq: BBdil2} . Examining these makes it clear that one needs to transform $\mathcal{V}_3$ such that it is mapped to $\mathcal{H}$. This may be achieved with a transformation in $v_3$ only
\beq
v_3\to v_3+\sqrt{2} \mathcal{Z}
\eeq
under this which
\beq
\mathcal{V}_3\to \mathcal{H},~~~\Lambda\to dv_3
\eeq
and so vielbeins of eq \ref{eq :BBDualVeil2} are mapped to those of eq \ref{eq :BBDualVeil}. The map on the RR sector also follows trivially whilst the NS 2-form of eq \ref{eq: BBB2} is mapped to that of eq \ref{eq: BBB22} up to an exact. 

So it is clear that one may ``cure'' the bad asymptotics and $g_{\rho v_3}$ mixing of section \ref{sec: BB no B2} either by a gauge transformation in the NS 2-from before dualisation, or by a local diffeomorpism on the dual coordinate $v_3$ after the duality procedure is performed.

\subsection{ Details of the Dual Baryonic Branch Structure}
All that remains to compete the elucidation of the baryonic dual is to give supplementary details to section \ref{G-structures} on the dynamical $SU(2)$ structure. Actually, the derivation of the structure is essentially the same as that of the dual of the wrapped D5 solution in section \ref{detailsD5}, so we will only focus on the differences here.

The 10-d MW Killing spinors of baryonic branch obey the same projection as the wrapped D5 spinors (see eq \ref{eq: projectors}). However, whilst the internal spinors are still parallel, they now differ by a point dependent phase $e^{i \zeta(r)}=\mathcal{C}+i \mathcal{S}$
\beq
\epsilon_{1}=e^{A}(\xi_{+}\otimes (e^{i\zeta(r)/2}\eta_+)+\xi_{-}\otimes (e^{-i\zeta(r)/2}\eta_-)),~~\epsilon_{2}= e^{A}(\xi_{+}\otimes(e^{-i\zeta(r)/2} \eta_+)+\xi_{-}\otimes (e^{i\zeta(r)/2}\eta_-))
\eeq
where the Minkowski warp factor is now $e^{2A}= \frac{e^{\Phi}}{\mathcal{C}}$. We now follow the steps illustrated between eqs \ref{eq: gammade} and \ref{eq:projj} such that the $SU(3)$-structure of the baryonic branch takes canonical form.

The dual 10-d Killing spinors are given as in eqs \ref{eq: D5dualKillingspinor},\ref{eq: D5Omega}, however the $\zeta^a$ entering into their definition are now given by
\beq\label{eq: zetaBBsual}
\zeta^1=\frac{2\sqrt{2}e^{-g-k-\phi}v_2\cos\psi}{\sqrt{\mathcal{C}}},~~~
\zeta^2=-\frac{2\sqrt{2}e^{-g-k-\phi}v_2
\sin\psi}{\sqrt{\mathcal{C}}},~~~\zeta^3=\frac{2\sqrt{2} e^{-2g-\phi}\mathcal{H}}{\sqrt{\mathcal{C}}}.
\eeq
The new spinor $\hat \epsilon_2$ is:
\beq
	\hat \epsilon_2 = \frac{e^{\Phi/2}}{\sqrt{\mathcal{C}}} \big( \z_+ \otimes  (e^{-i\zeta(r)/2}\hat\eta^2_-) + \z_- \otimes (e^{i\zeta(r)/2}\hat\eta^2_+) \big)
\eeq
where $\hat\eta^2_-$ is still given by eq \ref{eq: etadualD5}.

The dynamic $SU(2)$-structure supported by the dual baryonic branch solution may be expressed as 
\beq
	\begin{aligned}
		\Phi_+ &= \frac{e^{A}}{8} e^{-i v\wedge w} \big( k_{\|} e^{-ij} - i k_{\perp} \omega \big) \ , \\
		\Phi_- &= \frac{i e^{A}}{8} e^{i\zeta(r)} (v+i w)\wedge \big( k_{\perp} e^{-ij} + i k_{\|} \omega \big).
	\end{aligned} 
\eeq
The forms and functions entering into these expressions are given by
\beq
        \begin{aligned}
                &e^{2A} = \frac{e^{\Phi}}{\mathcal{C}} \\
                &e^{i\zeta(r)}=\mathcal{C}+i\mathcal{S}\\
                &k_{\|} = \frac{\sin \alpha}{\sqrt{1+\zeta.\zeta}} 
\qquad k_{\perp} = 
\sqrt{\frac{\cos^2 \alpha + \zeta.\zeta}{1+\zeta.\zeta}} \\
                &z = w - i \, v = \frac{1}{\sqrt{\cos^2 \alpha + \zeta.\zeta}} 
\big( \sqrt{\Delta} \tilde e^3 + \zeta_2 \sin\alpha 
\tilde e^\theta + i (\sqrt{\Delta} \tilde e^\r 
+ \zeta_2 \sin \alpha \tilde e^{\varphi}) \big) \\
                &j = \tilde e^{\r3} + \tilde e^{\varphi \theta} 
+ \tilde e^{21} - 
v \wedge w \\
                &\omega =  \frac{i}{\sqrt{\cos^2 \alpha + \zeta.\zeta}} 
\big( \sqrt{\Delta} 
(\tilde e^\varphi+ i \tilde e^\theta) - \zeta_2 \sin\alpha (\tilde e^\r 
+ i \tilde e^3) \big) \wedge (\tilde e^2 + i \tilde e^1),
        \end{aligned}
\label{su2structurezz}\eeq
with $\zeta^a$ defined by \ref{eq: zetaBBsual}. Specifically the vielbeins $\tilde e$ that the structure is expressed in terms of a rotation of those in eq \ref{eq: BBdualveildash}. First one preforms a rotation by $\alpha$
\beq
	\begin{aligned}
		\hat e^{\varphi} &= \cos \a e^\varphi + \sin \a e^{2'} \\
		\hat e^2 &= -\sin \a e^\varphi + \cos \a e^{2'} \\
		\hat e^a &= e^a \text{ for } a \neq \varphi, 2',
	\end{aligned}
\eeq
and then rotates these vielbeins to get $\tilde e = R \hat e$, where the matrix $R$ is given by eq \ref{eq:RmatixD5} with $\zeta^a$ by eq \ref{eq: zetaBBsual}.

\section{Details of the non-Abelian 
T-duality on D6 branes on $S^3$}
\label{app:EOM}
In this appendix, we study another background, similar to the one described in the main part of this paper. We want to start with a solution of D6-branes wrapping a three-sphere in type IIA supergravity, that preserves $\mathcal{N} = 1$ supersymmetry. We first describe such a solution, then we apply a non-Abelian T-duality to find a new type IIB supergravity solution. We study this transformation at the level of the geometric structure. We then take advantage of this example to make general statements on $\mathcal{N} = 1$ type IIB supergravity solutions.

\subsection{The type IIA solution} 

We are interested in finding a solution of D6-branes in type IIA supergravity. For that purpose, we start by considering eleven-dimensional supergravity. Because we only want D6-branes, the M-theory solution is a background with no fluxes. Such a solution is described in \cite{Brandhuber:2001kq} or \cite{Cvetic:2001kp} (we follow the notation of the latter).
The metric of the solution is:
\beq
	ds_{11}^2 = dx_{1,3}^2 + ds_7^2 \,,
\eeq
where the seven-dimensional internal space has the metric
\beq
	ds_7^2 = dr^2 + a^2 [(\S_1+g\s_1)^2 + (\S_2 + g\s_2)^2] + b^2(\s_1^2 + \s_2^2) + c^2 (\S_3 + g_3 \s_3)^2 + f^2 \s_3^2 \,,
\eeq
with $a, b, c, f, g, g_3$ all functions of the radial coordinate $r$. Here the $\S, \s$ are left-invariant SU(2) forms:
\beq
	\begin{aligned}
		\s_1 &= \cos \psi_1 + \sin \psi_1 \sin \theta d\varphi \,, & \S_1 &= \cos \psi_2 + \sin \psi_2 \sin \tilde\theta d\tilde\varphi \,, \\
		\s_2 &= -\sin \psi_1 + \cos \psi_1 \sin \theta d\varphi \,, & \S_1 &= -\sin \psi_2 + \cos \psi_2 \sin \tilde\theta d\tilde\varphi \,, \\
		\s_3 &= d\psi_1 + \cos \theta d\varphi \,, & \S_3 &= d\psi_2 + \cos \tilde\theta d\tilde\varphi \,. 
	\end{aligned}
\eeq
The BPS equations of this solution give \cite{Cvetic:2001ih}
\beq
	\begin{aligned}
		g &= -\frac{a f}{2b c} \,, & g_3 &= 2g^2 -1 \,,\\
		a' & = -\frac{c}{2a} + \frac{a^5 f^2}{8b^4 c^3} \,, & b' &= -\frac{c}{2b} - \frac{a^2(a^2-3c^2)f^2}{8b^3c^3} \,, \\
		c' &= -1 + \frac{c^2}{2a^2} + \frac{c^2}{2b^2} -\frac{3a^2f^2}{8b^4} \,, & f' &= -\frac{a^4f^3}{4b^4c^3} \,.
	\end{aligned}
\eeq
To get a ten-dimensional solution, we reduce the solution above along a U(1) isometry. To accomplish our goal of getting D6-branes wrapping a three-sphere, we choose the isometry generated by the Killing vector $\partial_{\psi_1} + \partial_{\psi_2}$. After some algebra, we get the following type IIA solution in string frame:
\beq
	\begin{aligned}
		&ds_{10}^2 = \alpha' g_s N e^{2A} \Big[ \frac{\mu}{\alpha' g_sN} dx_{1,3}^2 + dr^2 + b^2 (d\theta^2 + \sin^2 \theta d\varphi^2) + a^2 (\w_1+g d\theta)^2 \\
		&\qquad\qquad\qquad\quad+ a^2 (\w_2 + g \sin\theta d\varphi)^2 + h^2(\w_3 - \cos\theta d\varphi)^2 \Big] \,, \\
		&h^2 = \frac{c^2f^2}{f^2+c^2(1+g_3)^2} \,,\\
		&e^{4\F/3} = \frac{c^2 f^2}{4 (g_s N)^{2/3} h^2} \,,\\
		&e^{4A}=\frac{c^2 f^2}{4 h^2} \,, \\
		&\frac{F_2}{\sqrt{\alpha'} g_s N} =-(1+K)\sin\theta d\theta\wedge d\varphi +(K-1)\omega_1\wedge\omega_2-K'dr\wedge(\omega_3-\cos\theta d\varphi) \,,\\
		&K = \frac{f^2 - c^2(1-g_3^2)}{f^2+ c^2(1+g_3)^2}\,,
	\end{aligned}
\eeq
where the $\w$ are defined as $\S$, replacing $\psi_2$ with $\psi = \psi_2-\psi_1$.

\subsection{Non-Abelian T-dual}

Let us now take the solution from the previous section, and apply a non-Abelian T-duality on the SU(2) isometry parametrised by the $\w$. We follow Section 2 of \cite{Itsios:2013wd} and fix the gauge as $\tilde\theta = \tilde\varphi = v_1 = 0$. We obtain a type IIB supergravity solution. The metric, in string frame, is given by:
\beq
	\begin{aligned}
		ds_{IIB,st}^2&\!=e^{2A}\Bigg[dx_{1,3}^2 + N dr^2 + N b^2(d\theta^2 + \sin^2\theta d\varphi^2)\Bigg]+   
\frac{1}{\det{M}}\Bigg[2 (v_3 dv_2\!+\!v_3 dv_3)^2\!+\\   
&~~~\!4 a^2 e^{4 A}N^2\bigg(g^2 \left(a^2 v_2^2 (\hat{\omega}_2)^2\!   
+\!h^2 v_3^2 \left((\hat{\omega}_1)^2\!+\!(\hat{\omega}_2)^2\right)\right)\!-a^2 dv_3 (dv_3-2 g v_2 \hat{\omega}_2)+\\   
&~~~2 g h^2 v_3 \hat{\omega}_2   
   dv_2+h^2 dv_2^2-2 g h^2 v_2 v_3 \hat{\omega}_1 \hat{\omega}_3+h^2   
   v_2^2 (\hat{\omega}_3)^2\bigg)\Bigg] \,,
	\end{aligned}
\eeq
where   
\beq   
\det{M}=4 e^{2 A} \left(2 a^4 h^2 e^{4 A}+a^2 v_2^2+h^2   
   v_3^2\right)\,,   
\eeq   
and
\beq   
\hat{\omega}_1= \cos \psi\, d\theta - \sin \psi\, \sin \theta\, d\varphi,~~~\hat{\omega}_2= -\sin \psi\, d\theta - \cos \psi\, \sin \theta\, d\varphi,~~~\hat{\omega}_3= d\psi - \cos \theta d\varphi\,.   
\eeq
The dual dilaton $\hat \Phi$ is defined through   
\beq   
e^{-2\hat\Phi}=\det{M} e^{-2\Phi} \,,
\eeq   
and the two-form potential as  
\beq   
\begin{aligned}     
B_2&=-\cos\theta d\varphi   
\wedge dv_3\!+\! \frac{4\sqrt{2}a^4gh^2e^{6A}N^3}{\det{M}}\bigg(\hat{\omega}_1\wedge dv_2\!+\!(gv_3\hat{\omega}_1\!-\!v_2   
\hat{\omega}_3)\wedge\hat{\omega}_2\bigg)\\   
&~~~\frac{2\sqrt{2}v_2e^{2A}N}{\det{M}}\bigg(\hat{\omega}_3\wedge   
(h^2v_3dv_2-a^2v_2dv_3)+a^2g\hat{\omega}_1\wedge(v_2dv_2+v_3dv_3)\bigg) \,.\\   
\end{aligned}   
\eeq   
The RR sector has all possible fluxes turned on. $F_1$ and $F_5$ can be expressed as follows:
\beq   
\begin{aligned}   
F_1&= 2N \left(v_3 dr K'+(K-1) dv_3\right) \,,\\   
F_5&= -\frac{2 a^2 h U e^{6A }N^2}{b^2 \det{M}} \left(\sqrt{2} \det{M}   
   vol_4\wedge dr-4 N^2a^2 b^2 h v_2 \hat{\omega}_1\wedge \hat{\omega}_2\wedge   
   \hat{\omega}_3\wedge dv_2\wedge dv_3\right) \,,\\   
U&=g^2(K-1)-(K+1)\,.\\   
 \end{aligned}   
\eeq   
$F_3$ is considerably more complicated:
\beq   
\begin{aligned}   
F_3 & =\frac{\sqrt{2}N}{\det{M}} \Bigg[8N^3a^4 h^2 e^{6 A } \Bigg(v_3 \left(g^2 (K-1)+K+1\right) \hat{\omega}_1\wedge   
   \hat{\omega}_2\wedge dv_3+\\   
   \vspace{3mm}   
   &~~~~~~~~~~K' \bigg(g v_3 (\hat{\omega}_1\wedge dv_2\wedge   
   dr\!+\!g v_3 \hat{\omega}_1\wedge \hat{\omega}_2\wedge dr\!+\!v_2\hat{\omega}_2\wedge \hat{\omega}_3\wedge dr)-\\   
      \vspace{3mm}   
   &~~~~~~~~~~~v_2 \hat{\omega}_3\wedge dv_2\wedge dr\bigg)+g (K-1) (\hat{\omega}_1\wedge dv_2\wedge   
  dv_3+v_2 \hat{\omega}_2\wedge \hat{\omega}_3\wedge dv_3)\Bigg)+\\   
     \vspace{3mm}   
   &~~~~~~~~~~4 e^{2 A}N   
   \Bigg((K-1) v_2 \left(a^2 g v_2 \hat{\omega}_1\wedge dv_2\wedge dv_3+h^2   
   v_3 \hat{\omega}_3\wedge dv_2\wedge dv_3\right)+\\   
      \vspace{3mm}   
   &~~~~~~~~~~a^2 v_2 K' \bigg(g v_2   
   v_3 \hat{\omega}_1\wedge dv_2\wedge dr\!+\!g v_3^2 \hat{\omega}_1\wedge   
  dv_3\wedge dr-\\   
     \vspace{3mm}   
   &~~~~~~~~~~v_2 (v_2 \hat{\omega}_3\wedge dv_2\wedge   
   dr\!+\!v_3 \hat{\omega}_3\wedge dv_3\wedge dr)\bigg)+\\   
      \vspace{3mm}   
   &~~~~~~~~~~(K+1)   
   v_3 \left(a^2 v_2^2\!+\!h^2 v_3^2\right) \hat{\omega}_1\wedge \hat{\omega}_2\wedge   
  dv_3\Bigg)\!\!+\!\det{M} (K\!+\!1) v_2 \hat{\omega}_1\wedge \hat{\omega}_2\wedge   
  dv_2\Bigg]\\   
\end{aligned}   
\eeq

\subsection{Spinors and structure}

In this section, we follow the conventions of Andriot's thesis \cite{Andriot} for the SU(3)$\times$SU(3)-structure. We start from the solution before T-duality, which has an SU(3)-structure. This is type IIA supergravity so the spinors are of different chirality. The spinors of the original solution are:
\beq
	\begin{aligned}
		\epsilon_1 &= e^{\Phi/6} \big( \z_+ \otimes \eta_+ + \z_- \otimes \eta_- \big) \,, \\
		\epsilon_2 &= e^{\Phi/6} \big( \z_+ \otimes \eta_- + \z_- \otimes \eta_+ \big) \,.
	\end{aligned}
\eeq
They define the following SU(3)-structure:
\beq
	\begin{aligned}
		J &= e^{r3} + (\a e^2+\b e^\varphi) \wedge e^{\theta} + (\a e^\varphi - \b e^2) \wedge e^{1} \,,\\
		\Omega &= (e^r + i \, e^3) \wedge (\a e^2 + \b e^\varphi + i \, e^\theta) \wedge (\a e^\varphi - \b e^2 + i \, e^1) \,,
	\end{aligned}
\eeq
where
\beq   
\alpha(r)=\frac{a g}{\sqrt{b^2+a^2g^2}},~~~\beta(r)=\frac{b}{\sqrt{b^2+a^2g^2}},~~~\alpha^2+\beta^2=1\,,   
\eeq   
in terms of the vielbein basis:
\beq \label{eq:originalVielbeins}
	\begin{aligned}
	&e^{r}=e^{\Phi/3}dr\,,~~~e^{\theta}=e^{\Phi/3}b \, d\theta\,,~~~e^{\varphi}=e^{\Phi/3}b \sin\theta d\varphi\,,\\   
	&e^{1}=e^{\Phi/3}a(\omega_1+g \, d\theta)\,,~~~e^{2}=e^{\Phi/3} a(\omega_1+g\sin\theta d\varphi)\,,~~~e^{3}=e^{\Phi/3}h(\omega_3- \cos\theta d\varphi)\,.  
	\end{aligned} 
\eeq
Let us rotate this veilbein basis to put the structure in its canonical form.
\beq \label{eq:rotationVielbein}
	\begin{aligned}
		\hat e^{\varphi} &= \b e^\varphi + \a e^2 \,,\\
		\hat e^2 &= \a e^\varphi - \b e^2 \,,\\
		\hat e^a &= e^a \text{ for } a \neq \varphi, 2 \,.
	\end{aligned}
\eeq
It is a rotation since $\a^2 + \b^2 = 1$, but it reverses the orientation.
With respect to this new basis, the structure is expressed as:
\beq
	\begin{aligned}
		\hat J &= \hat e^{r3} + \hat e^{\varphi \theta} + \hat e^{21} \,,\\
		\hat \Omega &= (\hat e^r + i \, \hat e^3) \wedge (\hat e^\varphi + i \, \hat e^\theta) \wedge (\hat e^2 + i \, \hat e^1) \,.
	\end{aligned}
\eeq
That means that the spinors obey the following projections:
\beq \label{eq:proj}
	\hat \gamma_{\varphi \theta} \eta_+ = \hat \gamma_{r3} \eta_+ = \hat \gamma_{2 1} \eta_+ = i \, \eta_+ \,,
\eeq
where the $\hat \gamma$ matrices are defined in terms of the rotated vielbein basis.

Let us now look at the non-Abelian T-duality. We know that the spinors transform in the following way:
\beq
	\tilde \epsilon_1 = \epsilon_1 \, , \qquad \tilde \epsilon_2 = \Omega \, \epsilon_2 \,.
\eeq
$\Omega$ here is defined as:
\beq
	\Omega = \frac{\Gamma^{(10)}}{\sqrt{1 + \zeta.\zeta}} \big( -\Gamma^{123} + \zeta_1 \Gamma^1 + \zeta_2 \Gamma^2 + \zeta_3 \Gamma^3 \big) \,,
\eeq
where the $\zeta_a$ are given by
\beq
	\zeta_1 = -\frac{e^{-2\Phi/3} v_2 \cos \psi}{\sqrt{2} N a h}\,, \qquad \zeta_2 = \frac{e^{-2\Phi/3} v_2 \sin \psi}{\sqrt{2} N a h}\,, \qquad \zeta_3 = -\frac{e^{-2\Phi/3} v_3}{\sqrt{2} N a^2}\,.
\eeq
We are now going to consider the space after T-duality. The value for $\Omega$ above is written in the vielbein basis obtained directly from T-duality of the original basis \eqref{eq:originalVielbeins} without any rotation. To make things simpler, we are going to perform the same rotation with $\a, \b$ on this basis as before the T-duality (see \eqref{eq:rotationVielbein}), but we do not perform any rotation in $\psi$. We call this new basis $\check e$. It is defined in terms of the coordinate of the T-dual background as follows:
\beq
	\begin{aligned}
		&\check e^{r}=e^{\Phi/3}dr\,,~~~~~\check e^{\theta}=e^{\Phi/3}b \, d\theta\,,~~~~~\b \check e^{\varphi} + \a \check e^2 = e^{\Phi/3}b \sin\theta d\varphi \,,\\
		&\check e^{1}=\frac{2 \sqrt{N}e^{\phi/3} a}{\det M} \Big[ v_2(-\sqrt{2} v_3 \cos\psi + 2 e^{2\Phi/3} N a^2 \sin \psi) dv_3 \\
		&\qquad\qquad- (\sqrt{2} v_2^2 \cos\psi + 2 e^{2\Phi/3} N v_3 h^2 \sin \psi + 2 \sqrt{2} e^{4\Phi/3} N^2 a^2 h^2 \cos\psi) dv_2 \\
		&\qquad\qquad+ 2 e^{2\Phi/3} N g (-v_2^2 a^2 \sin\psi \hat\omega_2 + v_3 h^2(\sqrt{2} e^{2\Phi/3} N a^2 \sin\theta d\varphi + v_3 d\theta))\\
		&\qquad\qquad+ 2 e^{2\Phi/3} N v_2 h^2 (v_3 \cos\psi - \sqrt{2} e^{2\Phi/3} N a^2 \sin\psi) \hat\omega_3 \big]\,, \\
		&\a \check e^{\varphi} - \b \check e^{2}= \frac{2 \sqrt{N}e^{\phi/3} a}{\det M} \Big[ v_2(\sqrt{2}v_3 \sin\psi + 2 e^{2\Phi/3} N a^2 \cos \psi) dv_3\\
		&\qquad\qquad\quad+(\sqrt{2} v_2^2 \sin\psi - 2 e^{2\Phi/3} N v_3 h^2 \cos \psi+2 \sqrt{2} e^{4\Phi/3} N^2 a^2 h^2 \sin \psi) dv_2 \\
		&\qquad\qquad\quad +2 e^{2\Phi/3} N g ( - v_2^2 a^2 \cos\psi \hat\omega_2 + v_3 h^2 (-\sqrt{2} e^{2\Phi/3} N a^2  d\theta + v_3 \sin\theta d\varphi))\\
		&\qquad\qquad\quad + 2 e^{2\Phi/3} N v_2 h^2 (v_3 \sin\psi + \sqrt{2} e^{2\Phi/3} N a^2 \cos\psi) \hat\omega_3 \big] \,, \\
		&\check e^{3}=\frac{2\sqrt{N} e^{\Phi/3} h}{\det{M}} \big[ - \sqrt{2} v_2 v_3 dv_2 - \sqrt{2} (v_3^2 + 2 e^{4\Phi/3} N^2 a^4) dv_3 \\
		&\qquad\qquad\qquad+ 2 e^{2\Phi/3} N v_2 a^2(-v_3 g \hat\omega_1 + \sqrt{2} e^{2\Phi/3} N a^2 g \hat\omega_2 + v_2 \hat\omega_3) \big]\,.  
	\end{aligned}
\eeq
The projections obeyed by $\eta_+$ are still as in \eqref{eq:proj}. In this new basis, the T-dual $\Omega$ becomes:
\beq
	\Omega = \frac{1}{\sqrt{1 + \zeta.\zeta}} \big( - \alpha \check\Gamma^{1\varphi3} + \b \check\Gamma^{123} + \zeta_1 \check\Gamma^1 - \zeta_2 \b \check\Gamma^2 + \zeta_2 \alpha \check\Gamma^{\varphi} + \zeta_3 \check\Gamma^3 \big) \check\Gamma^{(10)} \,.
\eeq
So the new spinor $\tilde \epsilon_2$ is:
\beq
	\tilde \epsilon_2 = e^{\Phi/6} \big( \z_+ \otimes \check\eta^2_+ + \z_- \otimes \check\eta^2_- \big) \,,
\eeq
where
\beq\label{eq: etadualD5}
	\check\eta^2_+ = \frac{-\b \check\gamma^r - \zeta_1 \check \gamma^1 + \zeta_2 \b \check \gamma^2 - \zeta_3 \check\gamma^3 - \zeta_2 \alpha \check \gamma^\varphi}{\sqrt{1 + \zeta.\zeta}} \eta_- + i \frac{\alpha}{\sqrt{1 + \zeta.\zeta}} \eta_+ \,.
\eeq
It is clear here that, as long as $\alpha \neq 0$, we are in the general SU(3)$\times$SU(3)-structure case.
In order to simplify the expressions, we are performing a transformation of the vielbein basis:
\beq
	R = \frac{1}{\sqrt{\Delta}}\left(
	\begin{array}{cccccc}
		\b & 0 & 0 & \zeta_1 & -\zeta_2\b & \zeta_3 \\
		0 & \sqrt{\Delta} & 0 & 0 & 0 & 0 \\
		0 & 0 & \sqrt{\Delta} & 0 & 0 & 0 \\
		-\zeta_1 & 0 & 0 &\b & \zeta_3 & \zeta_2\b \\
		\zeta_2\b & 0 & 0 & -\zeta_3 & \b & \zeta_1 \\
		-\zeta_3 & 0 & 0 & -\zeta^2\b & -\zeta^1 & \b \\
	\end{array}
	\right)
\eeq
where
\beq
	\Delta = \b^2 +\zeta_1^2+\zeta_2^2 \b^2+\zeta^2_3
\eeq
We define a new basis:
\beq
	\tilde e = R.\check e
\eeq
In terms of this new basis, the spinor is:
\beq
	\tilde \eta^2_+ = -\left(\frac{\sqrt{\Delta} \tilde \gamma^r + \zeta_2 \alpha \tilde \gamma^\varphi}{\sqrt{1+\zeta.\zeta}} \right) \eta_- +i \frac{\alpha}{\sqrt{1 + \zeta.\zeta}} \eta_+
\eeq
And the projections in this basis are still:
\beq \label{eq:proj2}
	\tilde \gamma_{\varphi \theta} \eta_+ = \tilde \gamma_{r3} \eta_+ = \tilde \gamma_{2 1} \eta_+ = i \, \eta_+
\eeq

Let us now express the forms of the geometric structure, following the conventions of Andriot's thesis.
\beq
	\begin{aligned}
		&|a|^2 = e^{\Phi/3} \\
		&\theta_+ = \frac{\pi}{2} \qquad \theta_- = -\frac{\pi}{2} \\
		&k_{\|} = \frac{\alpha}{\sqrt{1+\zeta.\zeta}} \qquad k_{\perp} = \sqrt{\frac{\b^2 + \zeta.\zeta}{1+\zeta.\zeta}} \\
		&z = w - i \, v = \frac{1}{\sqrt{\b^2 + \zeta.\zeta}} \big( \sqrt{\Delta} \tilde e^r + \zeta_2 \alpha \tilde e^{\varphi}- i (\sqrt{\Delta} \tilde e^3 + \zeta_2 \alpha \tilde e^\theta) \big) \\
		&j = \tilde e^{r3} + \tilde e^{\varphi \theta} + \tilde e^{21} - v \wedge w \\
		&\omega =  \frac{-i}{\sqrt{\b^2 + \zeta.\zeta}} \big( \sqrt{\Delta} (\tilde e^\varphi+ i \tilde e^\theta) - \zeta_2 \alpha (\tilde e^r + i \tilde e^3) \big) \wedge (\tilde e^2 + i \tilde e^1)
	\end{aligned}
\eeq
In terms of those forms, the pure spinors are defined as:
\beq
	\begin{aligned}
		\Phi_+ &= \frac{|a|^2}{8} e^{i \theta_+} e^{-i v\wedge w} \big( k_{\|} e^{-ij} - i k_{\perp} \omega \big) \\
		\Phi_- &= \frac{i |a|^2}{8} e^{i\theta_-} (v+i w)\wedge \big( k_{\perp} e^{-ij} + i k_{\|} \omega \big)
	\end{aligned}
\eeq

Let us now look at the BPS equations of type IIB supergravity in the general case of SU(3)$\times$SU(3)-structure, generalising the system of pure SU(3)-structure that exhibit a rotation.

\section[BPS equations for a general IIB solution with SU(3)xSU(3)-structure]{BPS equations for a solution of type IIB supergravity with a general SU(3) $\times$ SU(3)-structure}
\label{app:XX}

We again follow the conventions of  Andriot's thesis in this section.
We start with the following pure spinors:
\beq
	\begin{aligned}
		\Phi_+ &= \frac{e^A}{8} e^{i \, \theta_+} e^{-i \, v\wedge w} (k_\| e^{-i\, j} - i \, k_{\perp} \w) \,,\\
		\Phi_- &= \frac{e^A}{8} e^{i \, \theta_-} (v + i \, w) \wedge (i \, k_\perp e^{-i\,j} - k_\| \w) \,.
	\end{aligned}
\eeq
For type IIB supergravity, the BPS equations are:
\beq
	\begin{aligned}
		(d - H\wedge) (e^{2A - \f} \Phi_-) &= 0 \,,\\
		(d - H\wedge) (e^{A - \f} \Re(\Phi_+)) &= 0 \,,\\
		(d - H\wedge) (e^{3A - \f} \Im(\Phi_+)) &= \frac{e^{4A}}{8} *_6 (F_1 - F_3 + F_5) \,.
	\end{aligned}
\eeq
Let us start with $\Phi_+$. We have:
\beq
	\begin{aligned}
		8 e^{-A} \Re(\Phi_+) &= k_\| \cos \theta_+ \left[ 1 + (\tan \theta_+ \chi  + \lambda) -\frac{1}{2} \left(\chi + \frac{1-\sin\theta_+}{\cos\theta_+} \lambda \right) \wedge  \left(\chi - \frac{1+\sin\theta_+}{\cos\theta_+} \lambda \right) \right] ,\\
		8 e^{-A} \Im(\Phi_+) &= k_\| \sin \theta_+ \left[ 1 - (\cot \theta_+ \chi - \lambda) - \frac{1}{2} \left(\chi+\frac{\cos\theta_+ + 1}{\sin\theta_+} \lambda \right) \wedge \left( \chi - \frac{\sin\theta_+}{\cos\theta_+ + 1} \lambda \right) \right] , 
	\end{aligned}
\eeq
where
\beq
	\begin{aligned}
		\chi &= j + v \wedge w + \frac{k_\perp}{k_\|} \Re(\w) \,,\\
		\lambda &= \frac{k_\perp}{k_\|} \Im(\w) \,.
	\end{aligned}
\eeq
Notice that, because of the various relations between the structure forms ($j \wedge \w = \w \wedge \w = 0$), we can use the following equations:
\beq
	\begin{aligned}
		&j \wedge \Re(\w) = j \wedge \Im(\w) = 0 \,,\\
		&\Re(\w) \wedge \Im(\w) = 0 \,,\\
		& \Re(\w) \wedge \Re(\w) = \Im(\w) \wedge \Im(\w) \,.
	\end{aligned}
\eeq
Using those, we can get the following relation:
\beq
	\lambda \wedge \lambda = k_\perp^2 \, \chi \wedge \chi \,.
\eeq
From there, we derive our first set of BPS equations. $(d - H\wedge) (e^{A - \f} \Re(\Phi_+)) = 0$ gives us
\beq
	\begin{aligned}
		&d[e^{2A - \f} k_\| \cos\theta_+] = 0 \,,\\
		&d[e^{2A - \f} k_\| \cos\theta_+ (\tan\theta_+ \chi + \lambda)] - e^{2A-\f} k_\| \cos \theta_+ H = 0 \,,\\
		&d\left[e^{2A - \f} k_\| \cos\theta_+ \left(\chi + \frac{1-\sin\theta_+}{\cos\theta_+} \lambda \right) \wedge  \left(\chi - \frac{1+\sin\theta_+}{\cos\theta_+} \lambda \right)\right] \\
		&\quad+ 2 e^{2A-\f} k_\| \cos \theta_+ H \wedge (\tan\theta_+ \chi + \lambda) = 0 \,.
	\end{aligned}
\eeq
From those, it is easy to see that $H = dB$ where:
\beq
	B = \tan\theta_+ \chi + \lambda \,,
\eeq
and the third equation simplifies into:
\beq
	d[ e^{4A-2\f} \chi \wedge \chi] = 0 \,.
\eeq
Let us now turn to $(d - H\wedge) (e^{3A - \f} \Im(\Phi_+)) = \frac{e^{4A}}{8} *_6 (F_1 - F_3 + F_5)$. We get
\beq
	\begin{aligned}
		&d[e^{4A-\f} k_\| \sin\theta_+] = e^{4A} *_6 F_5 \,,\\
		&d[e^{4A-\f} k_\| \sin\theta_+ (\cot\theta_+ \chi - \lambda)] + e^{4A-\f} k_\| \sin\theta_+ H = e^{4A} *_6 F_3 \,,\\
		&d\left[e^{4A-\f} k_\| \sin\theta_+ \left(\chi+\frac{\cos\theta_+ + 1}{\sin\theta_+} \lambda \right) \wedge \left( \chi - \frac{\sin\theta_+}{\cos\theta_+ + 1} \lambda \right)\right] \\
		&\qquad-2 e^{4A-\f} k_\| \sin\theta_+ H \wedge (\cot\theta_+ \chi - \lambda) = -2 e^{4A} *_6 F_1 \,.
	\end{aligned}
\eeq
Using all the equations we have so far, we can rewrite the three-form ones as:
\beq
	\begin{aligned}
		&H = d\lambda + \frac{e^\f \sin \theta_+}{k_\|} \left[ *_6 F_3 + (*_6 F_5)\wedge \lambda \right]  + \frac{e^\f \cos \theta_+}{k_\|} d(e^{-\f} k_\| \sin \theta_+) \wedge \chi \,,\\
		&e^{-2A} d(e^{2A} \chi) = \frac{e^\f \cos \theta_+}{k_\|} \left[ *_6 F_3 + (*_6 F_5)\wedge \lambda \right]  - \frac{e^\f \sin \theta_+}{k_\|} d(e^{-\f} k_\| \sin \theta_+) \wedge \chi \,.
	\end{aligned}
\eeq
Those equations have been written in such a way as to make the limits for $\theta_+ \rightarrow 0, \pi/2$ obvious, and to give the equations of the rotation present in \cite{Gaillard:2010qg} when taking $k_\perp \rightarrow 0, k_\| \rightarrow 1$ (limit of SU(3)-structure).
The last equation, involving $*_6 F_1$ can be rewritten in the following way:
\beq
	\frac{1}{2} d(e^{-\f} k_\| \sin \theta_+) \wedge \chi \wedge \chi = *_6 F_1 + (*_6 F_3) \wedge \lambda + (*_6 F_5) \wedge \lambda \wedge \lambda \,.
\eeq 
In summary, the BPS equations we get from $\Phi_+$ are:
\beq
	\begin{aligned}
		&d[e^{2A - \f} k_\| \cos\theta_+] = 0 \,,\\
		&d[e^{4A-\f} k_\| \sin\theta_+] = e^{4A} *_6 F_5 \,,\\
		&H = d\lambda + \frac{e^\f \sin \theta_+}{k_\|} \left[ *_6 F_3 + (*_6 F_5)\wedge \lambda \right]  + \frac{e^\f \cos \theta_+}{k_\|} d(e^{-\f} k_\| \sin \theta_+) \wedge \chi \,, \\
		&e^{-2A} d(e^{2A} \chi) = \frac{e^\f \cos \theta_+}{k_\|} \left[ *_6 F_3 + (*_6 F_5)\wedge \lambda \right]  - \frac{e^\f \sin \theta_+}{k_\|} d(e^{-\f} k_\| \sin \theta_+) \wedge \chi \,,\\
		&d[ e^{4A-2\f} \chi \wedge \chi] = 0 \,,\\
		&\frac{1}{2} d(e^{-\f} k_\| \sin \theta_+) \wedge \chi \wedge \chi = *_6 F_1 + (*_6 F_3) \wedge \lambda + (*_6 F_5) \wedge \lambda \wedge \lambda \,.
	\end{aligned}
\eeq
Let us now look at the equations we get for $\Phi_-$. We first define:
\beq
	\begin{aligned}
		\xi &= e^{i \, \theta_-} (v + i\, w) \,,\\
		\beta &= j - \frac{k_\|}{k_\perp} \omega \,.
	\end{aligned}
\eeq
We get for the BPS equations, after some simplifications:
\beq
	\begin{aligned}
		&d[e^{3A-\f} k_\perp \xi] = 0 \,,\\
		&k_\perp (d\beta + i \, H) \wedge \xi = 0 \,.
	\end{aligned}
\eeq 
The equation we would get for the six-form is just the one for the four-form wedged with $\beta$, so it is not an additional independent equation.

It is quite easy to check that, taking the pure SU(3) limit, that is $k_\| \rightarrow 0$, $k_\perp \rightarrow 1$, we recover the system we already knew from \cite{Gaillard:2010qg}.

Finally, we want to explicitly specialise to the cases of $\theta_+ = 0$ and $\theta_+ = \pi/2$.
First $\theta_+ = 0$:
\beq
	\begin{aligned}
		&d[e^{2A - \f} k_\|] = 0 \,,\\
		&F_5 = 0 \,,\\
		&H = d\lambda \,,\\
		&e^{-2A} d[e^{2A} \chi] = \frac{e^\f}{k_\|} *_6 F_3 \,,\\
		&d[ e^{4A-2\f} \chi \wedge \chi] = 0 \,,\\
		&*_6 F_1 + (*_6 F_3) \wedge \lambda = 0 \,,\\
		&d[e^{3A-\f} k_\perp \xi] = 0 \,,\\
		&k_\perp (d\beta + i \, H) \wedge \xi = 0 \,.
	\end{aligned}
\eeq
And, in the case $\theta_+ = \pi/2$:
\beq
	\begin{aligned}
		&d[e^{4A-\f} k_\|] = e^{4A} *_6 F_5 \,,\\
		&H = \frac{e^\f}{k_\|} *_6 F_3 + \frac{1}{e^{4A-\f} k_\|} d[e^{4A-\f} k_\| \lambda] \,,\\
		&d[e^{2A-\f} k_\| \chi] = 0 \,,\\
		&d[ e^{4A-2\f} \chi \wedge \chi] = 0 \,,\\
		&\frac{1}{2} d(e^{-\f} k_\|) \wedge \chi \wedge \chi = *_6 F_1 + (*_6 F_3) \wedge \lambda + (*_6 F_5) \wedge \lambda \wedge \lambda \,,\\
		&d[e^{3A-\f} k_\perp \xi] = 0 \,,\\
		&k_\perp (d\beta + i \, H) \wedge \xi = 0 \,.
	\end{aligned}
\eeq
Those systems do not look much more complicated than the ones in the pure SU(3) case, but there does not seem to be an easy transformation starting from either $\theta_+ = 0$ or $\theta_+ = \pi/2$ and recovering the full system.

%
%
%
%
%

  %
  %
  %
  %
  %
\end{document}